\documentclass{emulateapj}   

\usepackage{lineno}
\usepackage{rotating}
\usepackage{color}
\usepackage{hyperref}
\usepackage{breakurl} 

\def\ltsima{$\; \buildrel < \over \sim \;$}
\def\simlt{\lower.5ex\hbox{\ltsima}} 
\def\gtsima{$\; \buildrel > \over \sim \;$}
\def\simgt{\lower.5ex\hbox{\gtsima}} 

\def\deg{\hbox{$^\circ$}}

\def\phflux{photons\,cm$^{-2}$\,s$^{-1}$}
\def\gray{$\gamma$-ray}

\def\Halpha{H$_{\alpha}$}
\def\Hbeta{H$_{\beta}$}

\def\r95{$r_{\rm 95}$}
\def\p0{$\pi^{\rm 0}$}
\def\t0{$t_{\rm 0}$}
\def\ts{$t_{\rm s}$}

\def\Compton{\textit{Compton}}
\def\Fermi{\textit{Fermi}}

\def\Suzaku{\textit{Suzaku}}

\def\ncyg{{\rm V407 Cyg}}
\def\nsco{{\rm V1324 Sco}}
\def\nmon{{\rm V959 Mon}}
\def\ndel{{\rm V339 Del}}
\def\ncen{{\rm V1369 Cen}}
\def\nsgr{{\rm V5668 Sgr}}
\def\nvel{{\rm V382 Vel}}

\def\atel{{\rm ATel}}
\def\na{{\rm New A}}
\def\nar{{\rm New AR}}

\slugcomment{ApJ, accepted}

\shorttitle{\Fermi-LAT Gamma-ray Detections of Two Bright Classical Novae}
\shortauthors{Cheung et al.}

\begin{document}

\title{\Fermi\ LAT Gamma-ray Detections of Classical Novae V1369 Centauri 2013 and V5668 Sagittarii 2015}

\author{
C.~C.~Cheung\altaffilmark{1},
P.~Jean\altaffilmark{2,3},
S.~N.~Shore\altaffilmark{4,5}, 
\L.~Stawarz\altaffilmark{6}, 
R.~H.~D.~Corbet\altaffilmark{7,8},
J.~Kn\"odlseder\altaffilmark{2,3}, 
S.~Starrfield$^{9}$, 
D.~L.~Wood\altaffilmark{10},
R.~Desiante\altaffilmark{11,12}, 
F.~Longo\altaffilmark{13,14}, 
G.~Pivato\altaffilmark{15}, 
K.~S.~Wood\altaffilmark{10}
}

\altaffiltext{1}{Space Science Division, Naval Research Laboratory, Washington, DC 20375-5352, USA; Teddy.Cheung@nrl.navy.mil}
\altaffiltext{2}{CNRS, IRAP, F-31028 Toulouse cedex 4, France; Pierre.Jean@irap.omp.eu}
\altaffiltext{3}{GAHEC, Universit\'e de Toulouse, UPS-OMP, IRAP, Toulouse, France}
\altaffiltext{4}{Istituto Nazionale di Fisica Nucleare, Sezione di Pisa, I-56127 Pisa, Italy; steven.neil.shore@unipi.it}
\altaffiltext{5}{Dipartimento di Fisica ``Enrico Fermi", Universit\`a di Pisa, Pisa I-56127, Italy}
\altaffiltext{6}{Astronomical Observatory, Jagiellonian University, 30-244 Krak\'ow, Poland}
\altaffiltext{7}{Center for Research and Exploration in Space Science and Technology (CRESST) and NASA Goddard Space Flight Center, Greenbelt, MD 20771, USA}
\altaffiltext{8}{University of Maryland Baltimore County, Baltimore, MD 21250, USA}
\altaffiltext{9}{School of Earth and Space Exploration, Arizona State University, PO Box 871404, Tempe, AZ 85287-1404, USA}
\altaffiltext{10}{Praxis Inc., Alexandria, VA 22303, resident at Naval Research Laboratory, Washington, DC 20375-5352, USA}
\altaffiltext{11}{Universit\`a di Udine, I-33100 Udine, Italy}
\altaffiltext{12}{Istituto Nazionale di Fisica Nucleare, Sezione di Torino, I-10125 Torino, Italy}
\altaffiltext{13}{Istituto Nazionale di Fisica Nucleare, Sezione di Trieste, I-34127 Trieste, Italy}
\altaffiltext{14}{Dipartimento di Fisica, Universit\`a di Trieste, I-34127 Trieste, Italy}
\altaffiltext{15}{Istituto Nazionale di Fisica Nucleare, Sezione di Pisa, I-56127 Pisa, Italy}

\begin{abstract}

We report the \Fermi\ Large Area Telescope (LAT) detections of high-energy ($>$100\,MeV) \gray\ emission from two recent optically bright classical novae, V1369 Centauri 2013 and V5668 Sagittarii 2015. At early times, \Fermi\ target-of-opportunity observations prompted by their optical discoveries provided enhanced LAT exposure that enabled the detections of \gray\ onsets beginning $\sim$2\,days after their first optical peaks. Significant \gray\ emission was found extending to 39$-$55\,days after their initial LAT detections, with systematically fainter and longer duration emission compared to previous \gray\ detected classical novae. These novae were distinguished by multiple bright optical peaks that encompassed the timespans of the observed $\gamma$ rays. The \gray\ light curves and spectra of the two novae are presented along with representative hadronic and leptonic models, and comparisons to other novae detected by the LAT are discussed.

\end{abstract}



\begin{table*}
  \begin{center}
\caption{Summary of the two classical novae}
\begin{tabular}{lcc}
\hline
\hline
Name				   & \ncen\ 2013    & \nsgr\ 2015  \\
\hline
Distance$^{\rm a}$ (kpc) 	   & 2.5 	    & 2.0 \\
Discovery date$^{\rm b}$ 	   & 2013 Dec 2.41  & 2015 Mar 15.634 \\
                         	   & (MJD 56628.41) & (MJD 57096.634) \\
\hline
First optical peak mag$^{\rm c}$   & 3.6 	   & 4.1 \\
First optical peak date$^{\rm c}$  & 2013 Dec 5.7  & 2015 Mar 21.5 \\
                                   & (MJD 56631.7) & (MJD 57102.5) \\
Second optical peak mag$^{\rm c}$  & 3.4 	   & 4.4 \\
Second optical peak date$^{\rm c}$ & 2013 Dec 14.3 & 2015 Apr 3.7$-$6.4 \\
                                   & (MJD 56640.3) & (MJD 57115.7$-$57118.4) \\
\hline
Optical discovery name$^{\rm d}$ & PNV J13544700$-$5909080    & PNV J18365700$-$2855420 \\
Optical RA, Decl.$^{\rm d}$      & 208\fdg6958, $-$59\fdg1522 & 279\fdg2375, $-$28\fdg9283 \\
Optical $l, b$          	   & 310\fdg9847, +2\fdg7256    &   5\fdg3799, $-$9\fdg8668 \\
LAT RA, Decl.$^{\rm d}$ 	   & 208\fdg82, $-$59\fdg17    & 279\fdg17,  $-$29\fdg03 \\
LAT $95\%$ error radius 	   & 0\fdg10 & 0\fdg14 \\
Optical-LAT offset      	   & 0\fdg07  & 0\fdg12 \\
\hline
\end{tabular}
\end{center}
\smallskip
$^{\rm a}$ We adopted distances of $\sim$2.5\,kpc for \ncen\ \citep{sho14} and $\sim$2.0\,kpc for \nsgr\ \citep{ban15}.\\
$^{\rm b}$ The pre-discovery date is quoted for \ncen\ (see footnote 1 in the main text).\\
$^{\rm c}$ Based on the AAVSO data (see \S~1 and Figure~1). The first optical peak for \ncen\ was taken as an average of measurements taken on December 5.694 (3.4\,mag) and 5.708 (3.7\,mag); its second peak was based on an average of four measurements from December 14.273 (3.3\,mag), 14.292 (3.5\,mag), 14.302 (3.5\,mag) and 14.310 (3.27\,mag).  For \nsgr, we quote an observed first peak 4.1\,mag on March 21.501 because taking adjacent measurements around its actual observed peak 4.0\,mag on March 21.825 into account, March 21.789 (4.4\,mag), 21.796 (4.2\,mag), and 21.826 (4.3\,mag), indicated a fainter, 4.2\,mag average around March 21.8; its second peak is more broad, extending over $\sim$3\,days.\\
$^{\rm d}$ All positions are in the J2000.0 equinox.\\
\label{table-1}
\end{table*}

\section{Introduction\label{section-intro}}

Novae are thermonuclear runaway explosions triggered on the surfaces of white dwarf stars due to mass accretion from the companion in a binary system \citep{kra64}. In their nascent optically thick stage, a transient source of nuclear decay \gray\ emission with signature lines at $\sim$MeV energies will be produced by disintegration of unstable isotopes and positron annihilation \citep{cla74,sta78,cla81}, but no individual novae have yet been detected in this way \citep{her08}. Rather, novae have long been recognized visually by their telltale bright optical transient nature \citep{lun21,mcl39,pay57}, and contemporary radio, infrared, optical/ultraviolet, and X-ray imaging and spectroscopy studies have revealed different facets of the physics of the ensuing expanding ejecta \citep{bod08,wou14}.

The study of novae has undergone a recent resurgence because of the \Fermi\ Large Area Telescope \citep[LAT;][]{atw09} detections of $>$100\,MeV \gray\ emission from three classical novae \citep{ack14} following the initial LAT detection of the symbiotic-like recurrent nova \ncyg\ 2010 \citep{abd10,che10}. In hindsight, the idea that symbiotic novae can be continuum emitters in $>$100\,MeV $\gamma$ rays could have been expected from shock-accelerated particles in the ejecta through interactions with the dense wind of the red giant companion as considered for RS Oph \citep{tat07,her14}. However, the subsequent LAT detections of classical novae were truly unexpected, because they instead involve less-evolved (main sequence or subgiant) companions demonstrating high-energy particle acceleration occurring in the absence of dense wind material. Moreover, the LAT-detected classical novae displayed different optical/ultraviolet properties \citep[e.g.,][]{geh98} that indicated explosions from white dwarfs with an oxygen-neon (ONe) composition in \nmon\ 2012, while \nsco\ 2012 and \ndel\ 2013 appeared to be carbon-oxygen (CO) types.  The key implication of these handful of reported LAT detections in the local volume within $\sim$4$-$5\,kpc is that most novae, if not all, are potential sources of high-energy continuum \gray\ emission. 

Here, we report two additional classical novae detected with the \Fermi-LAT following their optical discoveries, the fifth and sixth novae now securely detected in $>$100\,MeV $\gamma$ rays. Both novae were discovered by J.~Seach, in Australia: V1369 Centauri (\ncen\ = Nova Cen 2013) on 2013 December 2.692\footnote{Note the report (\burl{http://www.aavso.org/possible-nova-centaurus}) that, ``Steven Graham in Templeton, NZ, has a webcam set up  ... hearing of this nova, he searched his archives and found it was visible on his image at 0945UT (Dec 2), but not on the previous dark one at 1330UT (Dec 1)."  This pre-discovery photo helps constrain the \ncen\ explosion epoch as 0.28\,days earlier than the Seach discovery (Table~1).}  and V5668 Sagittarii (\nsgr\ = Nova Sgr 2015 No.~2) on 2015 March 15.634 (all times are UT). Spectroscopic observations obtained on 2013 December 3 revealed \Halpha\ and \Hbeta\ emission lines and confirmed \ncen\ as a nova \citep[see][]{sea13} with a likely progenitor that brightened by $\sim$10\,mag \citep{kor13} at discovery. For \nsgr, Seach followed-up his 2015 March 15 discovery (6.0\,mag; it was not detected with a 10.5\,mag limit in his previous night's observation) with a bright \Halpha\ detection hours later \citep{sea15}, and the spectroscopic confirmation came the following day \citep{wil15}. 

\ncen\ and \nsgr\ were extensively monitored optically by the AAVSO and ARAS groups\footnote{American Association of Variable Star Observers (\burl{http://www.aavso.org/lcg}) and Astronomical Ring for Access to Spectroscopy (\burl{http://www.astrosurf.com/aras/}), respectively.} and showed broadly similar multi-peaked light curves. Their respective first peaks were 3.6\,mag and 4.1\,mag, making them the visually brightest novae observed since \nvel\ 1999 (2.5\,mag peak) and since the 2008 launch of \Fermi\ \citep[\ndel\ 2013 peaked at 4.5\,mag; see][]{mun15}. \nsgr\ was characterized by systematically longer timescales in its brightening and fading (see Figure~1 for an overview, and Table~1). Both rose to their first peaks by $\sim$1.9\,mag after discovery, but with different rates of $\sim$0.6\,mag\,day$^{-1}$ for \ncen\ and $\sim$0.3\,mag\,day$^{-1}$ for \nsgr. Following a 2-day decline, a slightly brighter second peak in \ncen\ was observed $\sim$9\,days after its first peak with three subsequent fainter peaks spaced by 5$-$14\,days. \nsgr\ declined over a longer 3.5-day interval and rebrightened to a modestly fainter, broader second peak $\sim$15$-$18\,days after its first peak, and underwent three additional brightening events to $\sim$4.5$-$5\,mag at intervals of 8$-$13\,days \citep[see][]{waa15}.

In the following, we present the LAT observations of the two novae (\S~2), describing the \gray\ light curves coinciding with multi-peaked optical activity \citep[\S~2.1; preliminary results were presented in][]{che15c}, and hadronic and leptonic modeling of their high-energy \gray\ spectra (\S~2.2). The results are discussed in \S~3, and we conclude in \S~4.

\begin{figure*}[t]
  \begin{center}
   \includegraphics[width=8cm]{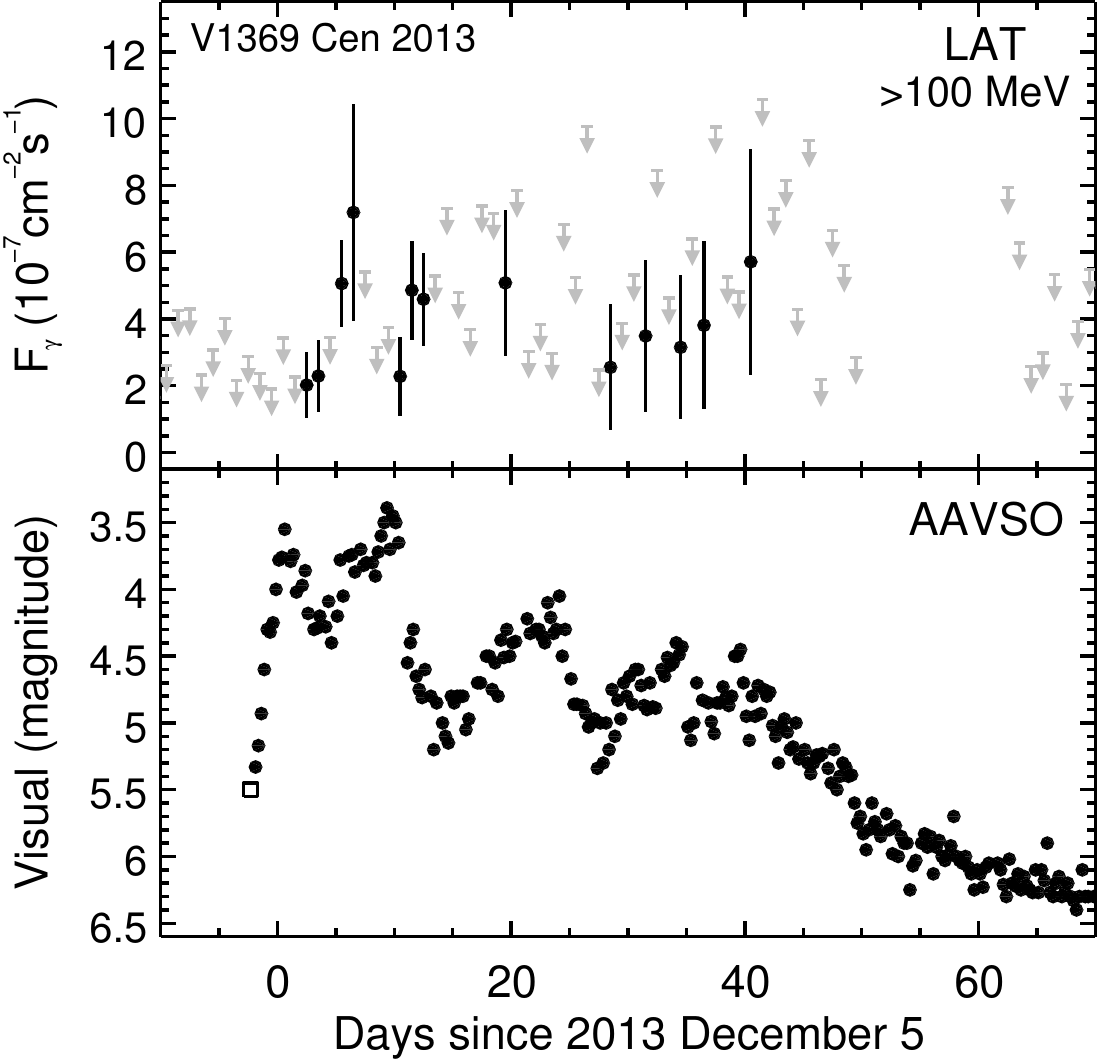}\hspace{.25cm}\includegraphics[width=8cm]{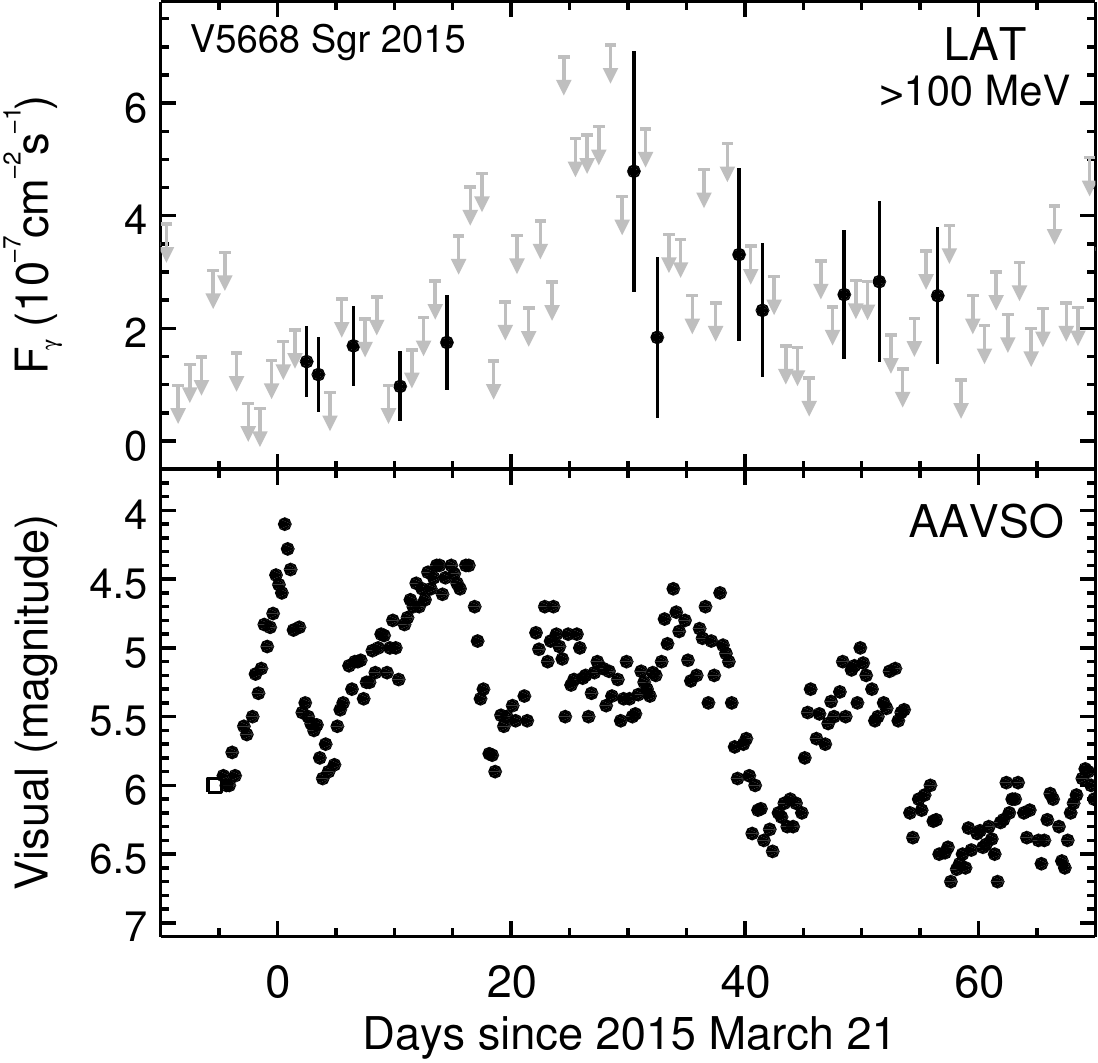}
  \end{center}
\caption{\Fermi-LAT 1-day bin $>$100\,MeV \gray\ (top panels) and optical (bottom panels) light curves for \ncen\ (left) and \nsgr\ (right). For the LAT data, vertical lines indicate $1\sigma$ uncertainties when TS $\geq 4.0$ and gray arrows indicate $2\sigma$ upper limits when TS $<4.0$ (see Appendix A for tabulated fluxes, TS values, as well as corresponding exposures). The optical discovery measurements (open squares) are shown separately from the subsequent 0.25-day bin averaged data extracted from the AAVSO database (filled circles).}
\label{figure-1}
\end{figure*}

\section{LAT Observations and Analysis}

The \Fermi-LAT routinely operates in a sky-survey mode, providing coverage of the entire sky every $\sim$3.2 hrs (two spacecraft orbits). We obtained \Fermi\ target-of-opportunity (ToO) observations a few days after each optical nova discovery that resulted in exposures approximately twice what the nominal sky survey would have achieved in both cases. The ToOs were timed for after the optical emission peaked, as previous observations of \ndel\ 2013 indicated the $>$100\,MeV \gray\ onset occurred $\sim$1.5\,days later. This strategy led to the successful early detections of the two new novae, while continued \Fermi\ sky-survey observations at later times revealed detectable \gray\ emission out to $\sim$1.5$-$2\,months.

At the time of the \ncen\ optical discovery, \Fermi\ was operating with a modified sky survey pattern favoring the Galactic center direction\footnote{\burl{http://fermi.gsfc.nasa.gov/ssc/observations/types/exposure/}} with minimal sky exposure toward Centaurus. This \Fermi\ ToO observation began on 2013 December 6.0, just 0.3\,day after \ncen's first optical peak. The initial 5-day ToO sequence resulted in a low-significance cumulative 3-day LAT detection beginning about 1.5\,days after its optical peak \citep{che13a}. On the last day of this ToO sequence, the source brightened in $\gamma$ rays as the optical was rising to its second peak \citep{che13b}. This prompted a return to the ToO pointing for five additional days, providing enhanced coverage of the second, brighter, optical peak as well, after three days of modified sky survey. 

For \nsgr, the \Fermi\ TOO started around March 17.9, while the optical brightness continued to rise after discovery, and extended for 18\,days in total, ending on April 5; the ToO observation was longer because of the slower rise to optical peak. As the optical brightness climbed, there were no significant LAT detections until the expected onset in $\gamma$-rays $\sim$1.5\,days after optical peak \citep{che15a,che15b}, similar to that in \ncen\ and in the previous \ndel\ case.

For the subsequent LAT analysis, we selected 0.1$-$100\,GeV data with {\tt P7REP$\_$SOURCE$\_$V15} instrument response functions within a region-of-interest (ROI) centered on each optical nova position with a 15\deg\ radius. To model the background, we included all 3FGL catalog sources \citep{ace15} within each ROI as well as the diffuse \gray\ emission represented by the Galactic and isotropic templates\footnote{Respective files, gll$\_$iem$\_$v05$\_$rev1.fit and iso$\_$source$\_$v05.txt, available from \burl{http://fermi.gsfc.nasa.gov/ssc/data/access/lat/BackgroundModels.html}}. Data analysis was performed using \Fermi\ Science tools version 09-34-01. For the mix of data obtained as part of sky-survey and ToO pointings, data were selected following option 3 under the FSSC recommendations\footnote{\burl{http://fermi.gsfc.nasa.gov/ssc/data/analysis/documentation/Cicerone/Cicerone_Likelihood/Exposure.html}}. We used binned likelihood analysis with the {\tt gtlike} tool for the light curves and for the average spectral results. Systematic uncertainties can be estimated as $\sim$8$\%$ for the fluxes and $\sim$0.1 for the spectral slopes \citep{ack12}.

\begin{figure*}[t]
  \begin{center}
\includegraphics[width=7cm]{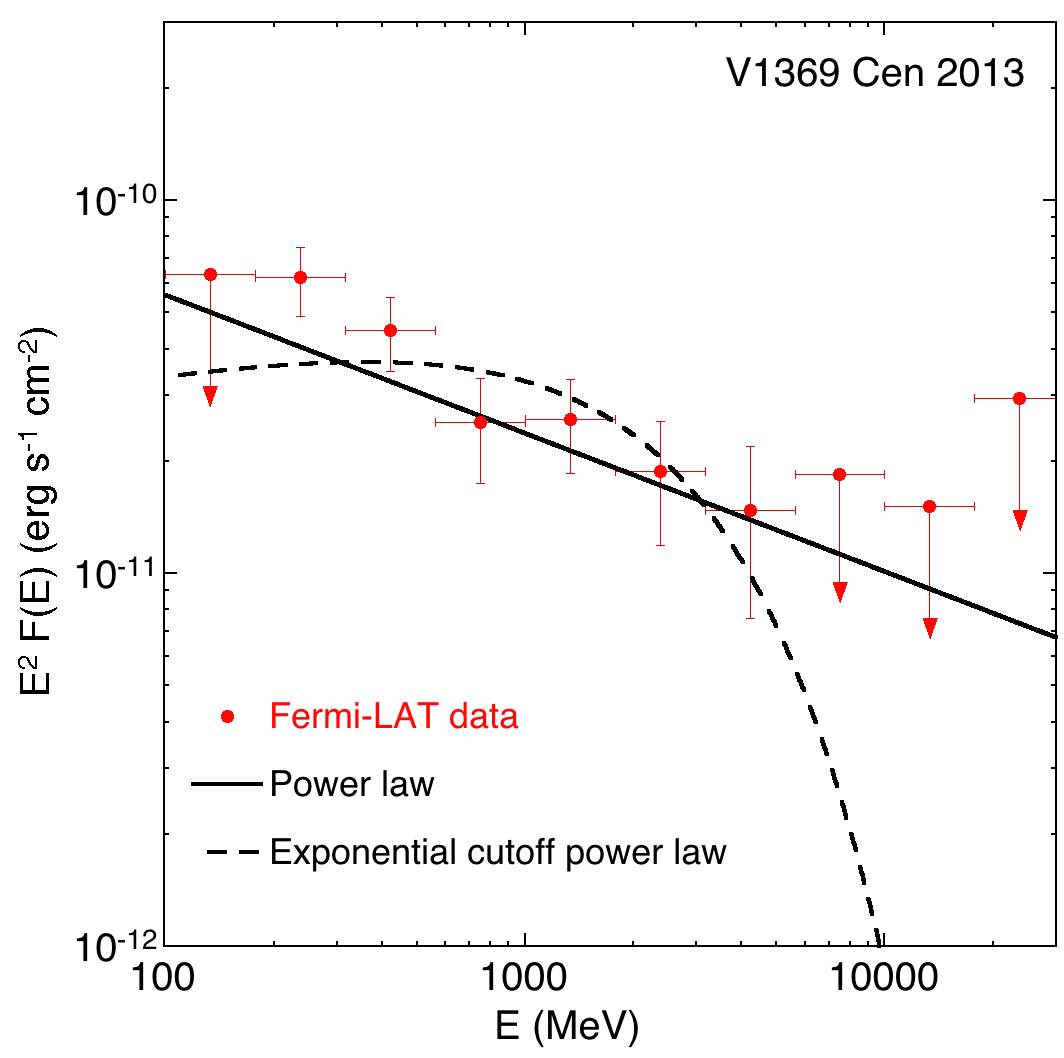}\includegraphics[width=7cm]{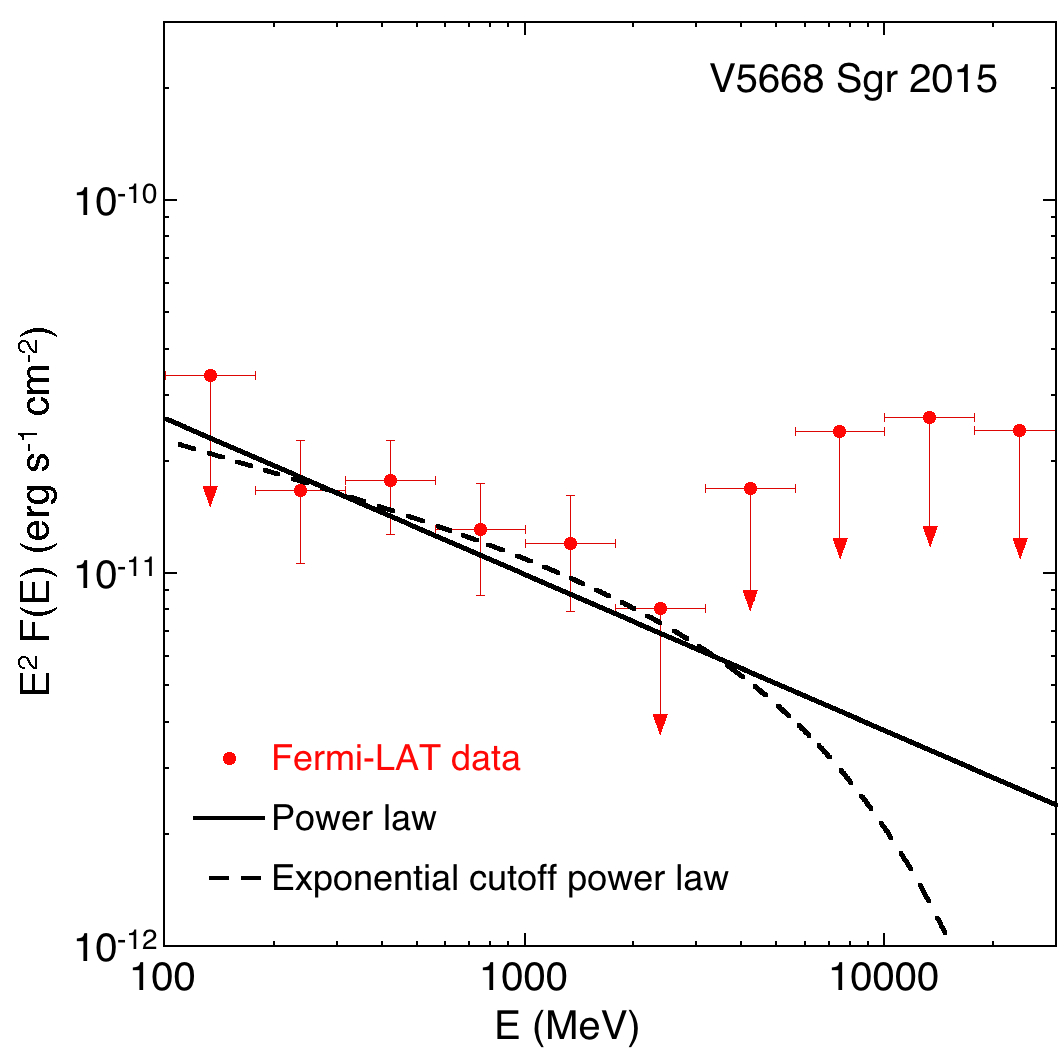}
\includegraphics[width=7cm]{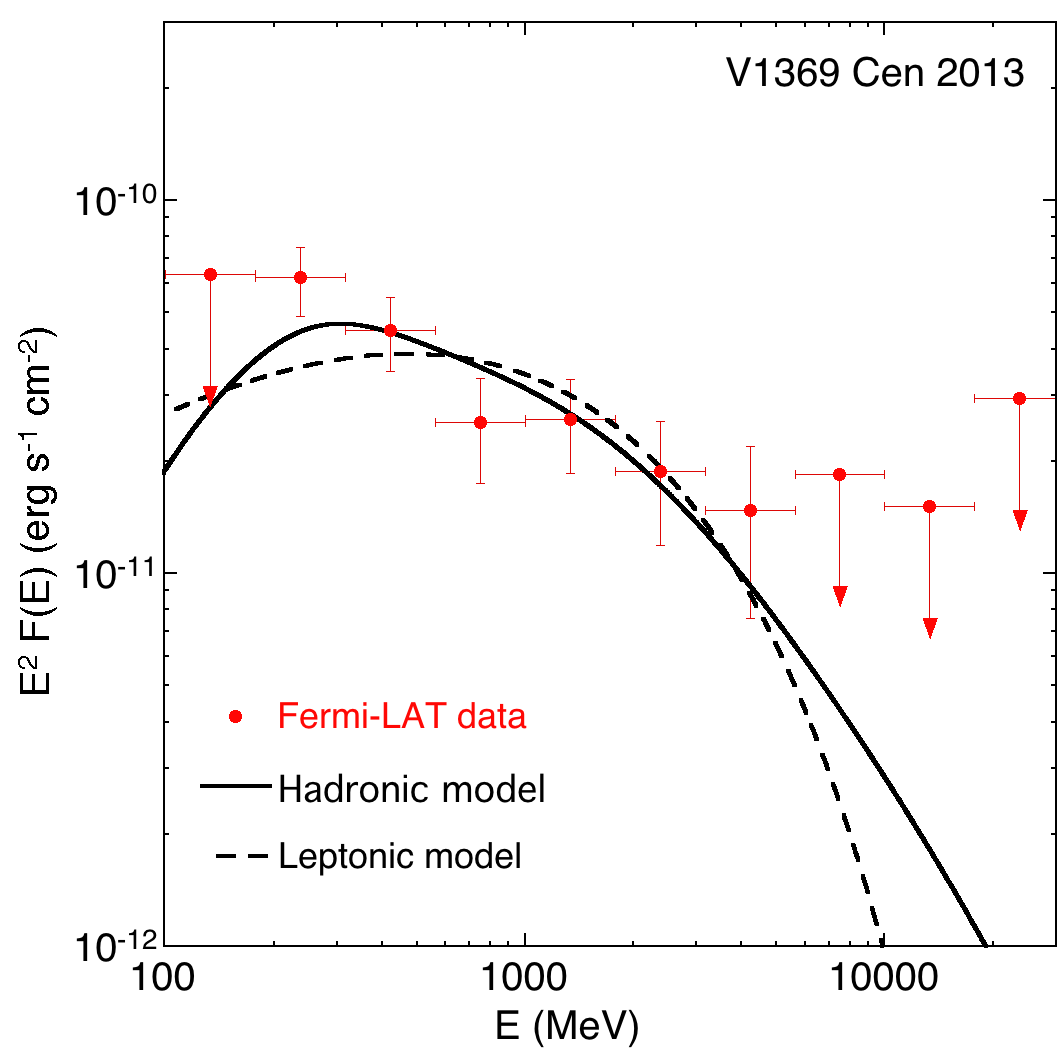}\includegraphics[width=7cm]{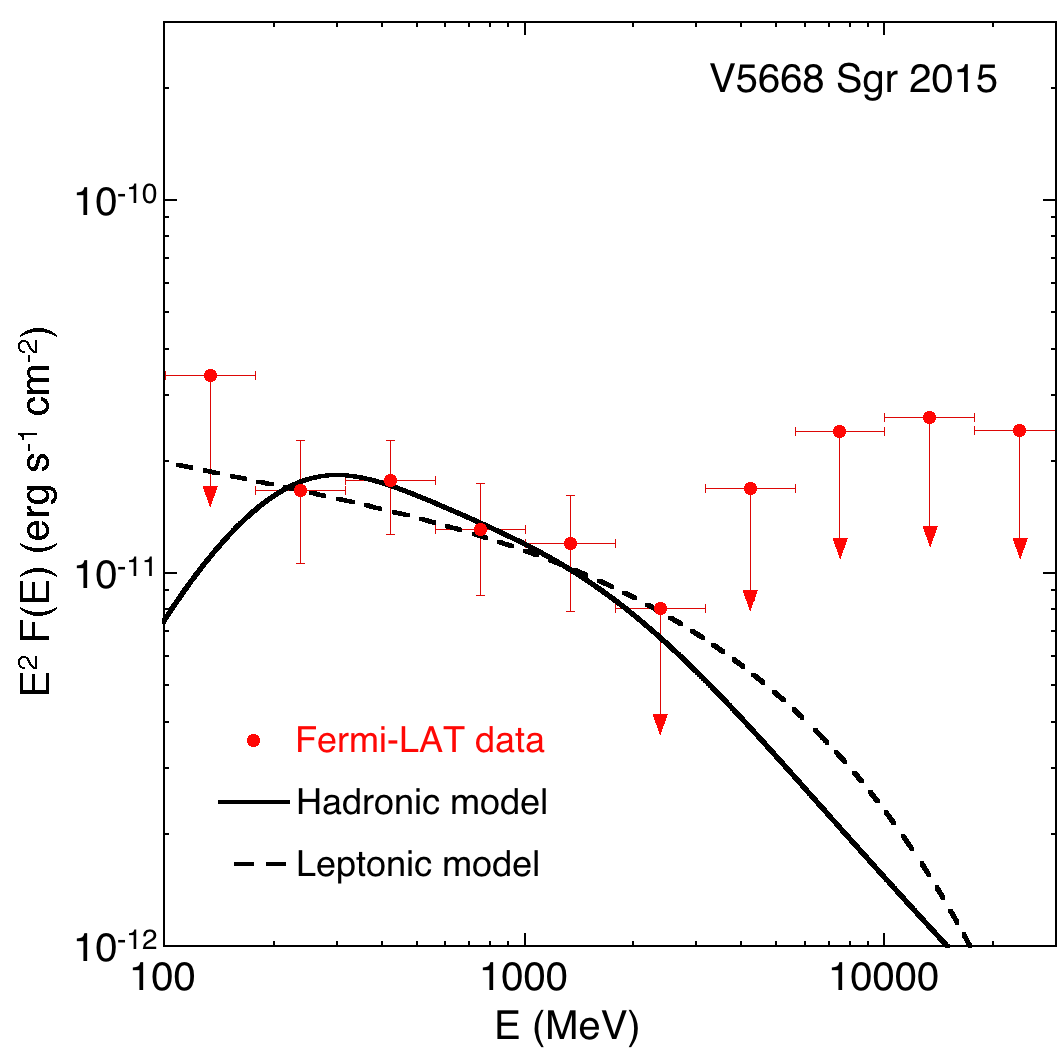}
  \end{center}
\caption{\Fermi-LAT $>$100\,MeV average spectra of \ncen\ (left panels) and \nsgr\ (right panels) with vertical bars indicating $1\sigma$ uncertainties for data points with TS $\geq 4$, otherwise, $2\sigma$ upper limits are plotted. Overlaid are the PL and ECPL fits (top panels) and hadronic and leptonic model fits (bottom panels).}
\label{figure-2}
\end{figure*}

\subsection{Light Curves\label{sec-lightcurve}}

For both novae, we analyzed 212\,days of data (four orbital precession periods of the \Fermi\ satellite) centered around starting times defined as the day of their first optical peaks, \ts\ = 2013 December 5.0 for \ncen\ and \ts\ = 2015 March 21.0 for \nsgr\ (see Table~1). Note our definition of \ts\ here is different from \citet{ack14} where the starting times were rather defined as the first significant 1-day bin LAT detection. The present choice is to facilitate comparison with \ndel\ 2013.  For \ncen, the data interval spanned \ts $-$106\,days to \ts $+$106\,days, but for \nsgr, the data analyzed were up to 2015 Jun 1.0 (MJD 57174.0), i.e., from \ts $-$140\,days to \ts $+$72\,days\footnote{Subsequently, we further extended the light curve by 24\,days, i.e., up to June 25 (\ts $+$96\,days) and the two most significant days were isolated points with test statistic, TS = 3.7 (June 5) and 3.8 (June 9) during an interval of low exposure. For one degree of freedom in the fits (the normalization in this case), the source significances are $\sim$$({\rm TS})^{1/2}$ \citep{mat96}, thus these additional $<2\sigma$ points do not add useful information.}. The long intervals analyzed with integer multiples of the \Fermi\ precession period were chosen to gauge the effects of variable LAT exposure on the signal determinations as well as the level of potential background contamination at the nova positions by inspecting the data before the optical explosions. 

The full 212-day datasets were used to determine the normalizations of known variable \gray\ sources in each nova field, as well as the Galactic and isotropic normalizations. For the field source analysis, we considered objects within 5\deg\ of each nova flagged as variable in the 3FGL catalog, as well as any potentially contaminating blazar-type sources within the 15\deg\ ROI. Only two sources satisfied the former criteria, with offsets of 4\fdg6 (3FGL J1328.9$-$5607) and 3\fdg3 (3FGL J1400.7$-$5605) from \ncen, and were found to be quite faint during this interval\footnote{See up-to-date LAT aperture light curves for all the 3FGL sources: \burl{http://fermi.gsfc.nasa.gov/ssc/data/access/lat/4yr_catalog/ap_lcs.php}}. Three highly-variable, bright blazars are found in the vicinity of \nsgr, but with large offsets of 7\fdg9 (3FGL J1833.6$-$2103), 11\fdg8 (3FGL J1917.2$-$1921), and 13\fdg3 (3FGL J1911.2$-$2006).

Assuming a point source at the position of each nova with LAT spectra parameterized as a single power law (PL), we generated 1-day bin $>$100\,MeV light curves with the photon index, $\Gamma = 2.2$, fixed to the the typical value measured in the previous LAT-detected classical novae \citep{ack14}. We also allowed the PL slope to be left free to fit, but the resultant indices varied widely and appeared unreliable because of the limited statistics in these short time-bins; thus the $\Gamma = 2.2$ results are presented. All the background parameters were fixed except for the normalizations of the Galactic diffuse component and the closest background blazar, 3FGL\,J1833.6$-$2103 = PKS~1830$-$211. Despite the enhanced LAT-ToO exposures at the early stages, the cumulative significances of \ncen\ and \nsgr\ (Table~2) are lower than the previous cases. This made it difficult to derive good constraints on the curvature of the spectra and temporal variations. Note the wide range of upper limits in the light curves due to the combination of sky survey and ToO exposure (see Appendix A). Also, there is a 12-day gap in the \ncen\ observations starting at \ts $+$50\,days due to a LAT ToO observation of SN2014J in M82. 

In the \gray\ and optical comparison (Figure~\ref{figure-1}),  we display the interval \ts\ $-$10\,days to +70\,days, with the AAVSO data averaged in 0.25\,day bins. In the 1-day bin LAT light curves (see Appendix A for corresponding TS and exposure values), the first significant \gray\ detections (defined as TS $>4$) of both novae spanned the two days from \ts\ $+$2.0 to $+$4.0\,days with TS $\sim$6$-$7 in each day for \ncen\ and TS = 11.5 and 5.3 for \nsgr, confirming the preliminary findings \citep{che13a,che13b,che15a,che15b}.  The respective $>$100\,MeV fluxes were $\sim$(2.0$-$2.3)\,$\times 10^{-7}$ and  $\sim$(1.3$-$1.6)\,$\times 10^{-7}$\,\phflux, both fainter than previously detected novae at similar times. By splitting the data for the first significant 1-day detections into two halves, we found the more significant detection in the latter half-day in \ncen\ (TS $= 1.1$ versus 6.0) and in the first half-day in \nsgr\ (TS $= 9.2$ versus 3.5). Taking these 0.5\,day-bin results, we can constrain the \gray\ onsets to $\sim$1.8 and 2.0\,days after their respective first optical peaks. The LAT ToOs for each nova provided enhanced exposure through the observed \gray\ onsets, suggesting the \gray\ onset determinations were not exposure artifacts. 

The total LAT-detected durations of these two novae are systematically longer than for the previously detected novae, albeit with the detections appearing sporadic because of a combination of the changing LAT exposure and low photon statistics. This limits our statements regarding changes in the \gray\ light curves predominantly to intervals with good exposure and precludes a meaningful comparison with the optical data when the LAT exposures were low. Considering primarily the 1-day bins with enhanced ToO exposures at early times (i.e., \ts\ to \ts\ $+$15\,days), we found only 3/10 days in \ncen\ with highly significant detections (TS = 19.3$-$37.6) and 3/15 days in \nsgr\ with significant detections (TS = 8.5$-$11.6), with no obvious correlation of TS with exposure. In the three 1-day bins with largest TS in \ncen, the fluxes are a factor of two greater than those in other comparably exposed time bins. These higher \gray\ fluxes were observed both preceding and following its second optical peak at \ts\ $+$9.3\,days (the more limited statistics in \nsgr\ prevent a similar determination). At later times (after \ts $+$15\,days), both novae declined in and out of detectability because of lower exposures when the LAT returned to survey mode.  In \ncen, there is only a single TS $>9$ point on \ts\ $+$19.5\,days, which amounts to a total detected duration of at least 18\,days for this nova (note the total detected durations include the first significant detections for each nova at \ts\ $+$2.0\,days), and a total detected duration of 39\,days is implied if the lower-significance (TS $>4$) detections found in five of the 21 subsequent bins up to \ts\ $+$40.5\,days are considered.  \nsgr\ was sporadically detected at late times in seven more bins evenly spread over a long interval up to +57.0\,days, again, without an apparent correlation with exposure. Its apparent lull in signal (essentially TS = 0) from \ts\ $+$15.0 to $+$24\,days is difficult to gauge because of minimal exposure at these times. If we take the last day with observed TS $>9$ in \nsgr, then the total detected duration is 47\,days, and this duration is enlarged slightly to 55\,days if we consider the single TS = 8.6 point at \ts\ $+$56.5\,days.

\begin{figure*}
  \begin{center}
\includegraphics[width=8cm]{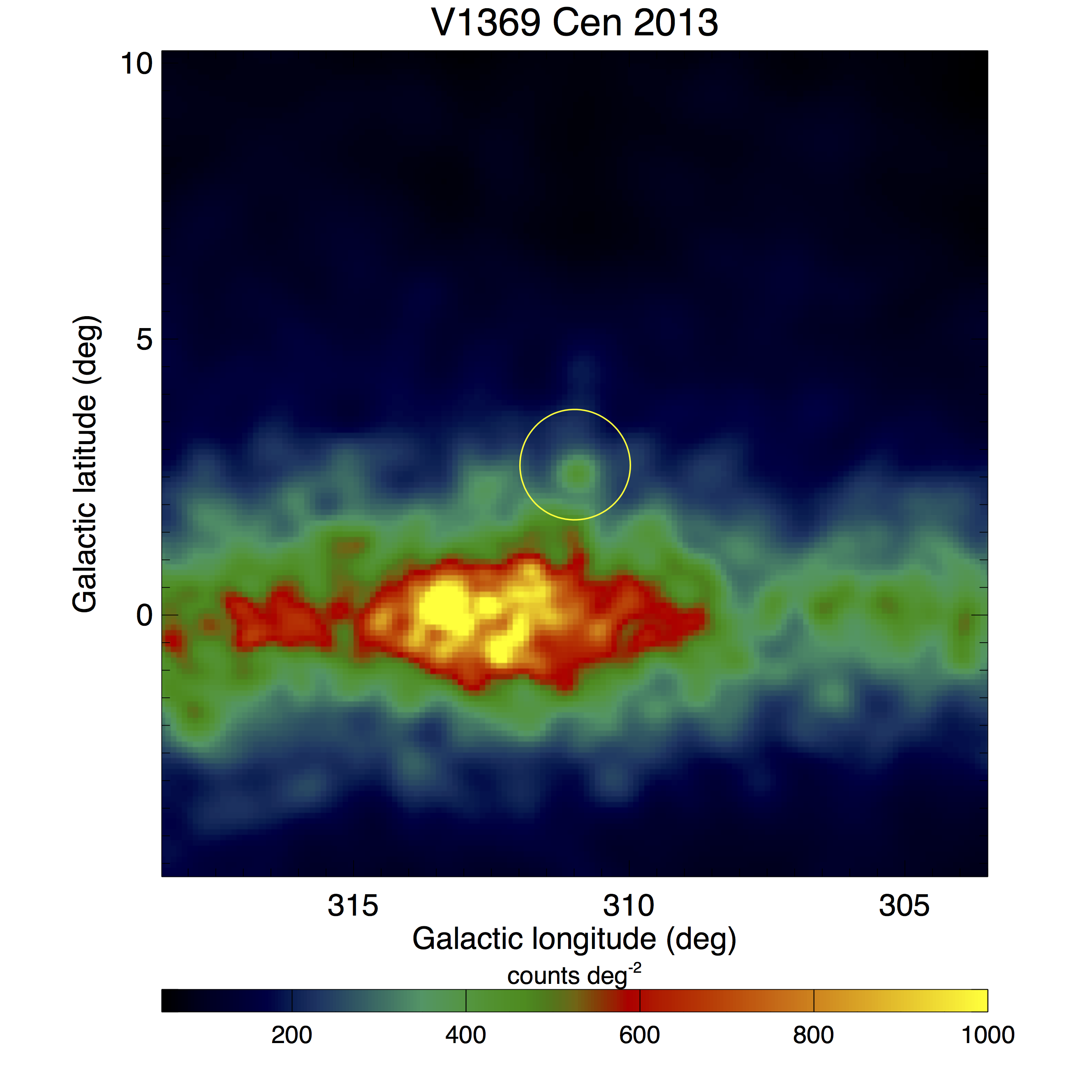}\hspace{-0.5cm}\includegraphics[width=8cm]{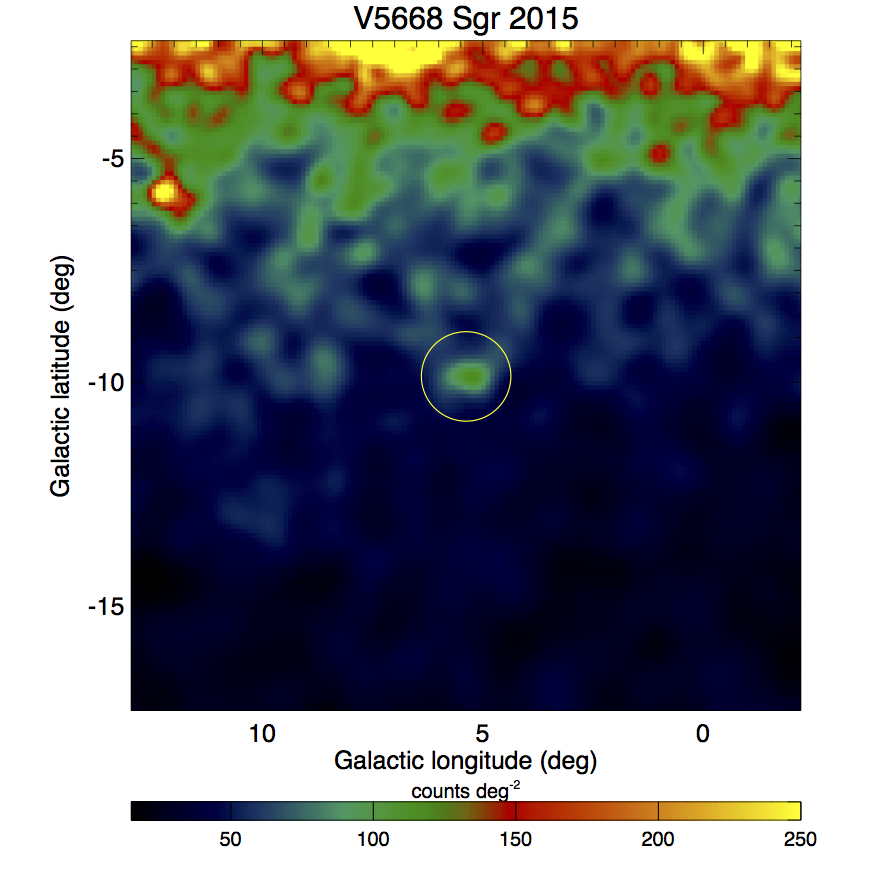}
  \end{center}
\caption{\Fermi-LAT \gray\ adaptively smoothed count maps centered on the optical positions of the two classical novae (marked with yellow 1\deg\ radius circles). The maps used $0\fdg1 \times 0\fdg1$ pixels with 39\,days of $>$100\,MeV data for \ncen\ (left) and 55\,days of $>$300\,MeV data for \nsgr\ (right). The slightly higher energy threshold for \nsgr\ was chosen for this figure alone in order to help suppress the visual prominence of the surrounding steep-spectrum diffuse Galactic emission.}
\label{figure-3}
\end{figure*}

\begin{table*}
  \begin{center}
\caption{\Fermi-LAT $>$100\,MeV \gray\ spectral and model fit results for the two classical novae}
    \tabcolsep 4.0pt
\begin{tabular}{lcc}
\hline\hline
\multicolumn{1}{l}{Nova} &
\multicolumn{1}{c}{\ncen\ 2013} &
\multicolumn{1}{c}{\nsgr\ 2015} \\
\hline
\hline
\multicolumn{3}{c}{Single Power Law} \\
\hline
Flux, $F_{\gamma}$ & $25.4 \pm 4.1$	& $11.4 \pm 2.2$ \\
Photon index            & $2.37 \pm 0.09$ & $2.42 \pm 0.13$ \\
TS           	               & 99.9           	& 70.4    \\
\hline
\hline
\multicolumn{3}{c}{Exponentially Cutoff Power Law} \\
\hline
Flux, $F_{\gamma}$ & $21.1 \pm 4.2$ 	& $10.9 \pm 2.4$  \\
Slope, $s$                & $1.84 \pm 0.23$	& $2.27 \pm 0.25$ \\
$E_{\rm c}$              & $2.3 \pm 1.0$  	& $9 \pm 14$ \\
TS           	               & 110.6 		& 71.1 \\
\hline
\hline
\multicolumn{3}{c}{Hadronic Model} \\
\hline
Flux $F_{\gamma}$ & $20.7 \pm 3.6$     & $8.2_{-1.4}^{+1.2}$ \\
Slope, $s_{\rm p}$   & $3.0_{-0.4}^{+0.9}$  & $3.1_{-0.9}^{+0.3}$ \\
$E_{\rm cp}$	      & $>$10 		  & $>$10 \\
TS  		              & 115.3 		  & 70.7 \\
\hline
\hline
\multicolumn{3}{c}{Leptonic Model} \\
\hline
Flux, $F_{\gamma}$ & $20.0^{+2.6}_{-4.4}$  & $10.1_{-2.2}^{+3.0}$ \\
Slope, $s_{\rm e}$ & $1.3^{+1.3}_{-1.6}$  & $2.8_{-1.5}^{+0.5}$ \\
$E_{\rm ce}$	    & $3.2^{+7.1}_{-1.4}$  & $>$3.2 \\
TS  		             & 111.2 		  & 71.1 \\
\hline
\end{tabular}
\normalsize
  \end{center} 
{\bf Notes.} 
Best-fit parameters from the average spectral fits and the hadronic and leptonic modeling of the LAT data for \ncen\ and \nsgr. The reported errors on spectral parameters are $1\sigma$ statistical uncertainties only with $>$100\,MeV fluxes, $F_{\gamma}$, in units of $10^{-8}$\,\phflux\ and the cutoff energies, $E_{\rm c}$, in GeV.
\label{table-2}
\end{table*}

\begin{figure*}[t]
  \begin{center}
\hspace*{0.185in}\includegraphics[width=6.1cm,angle=90]{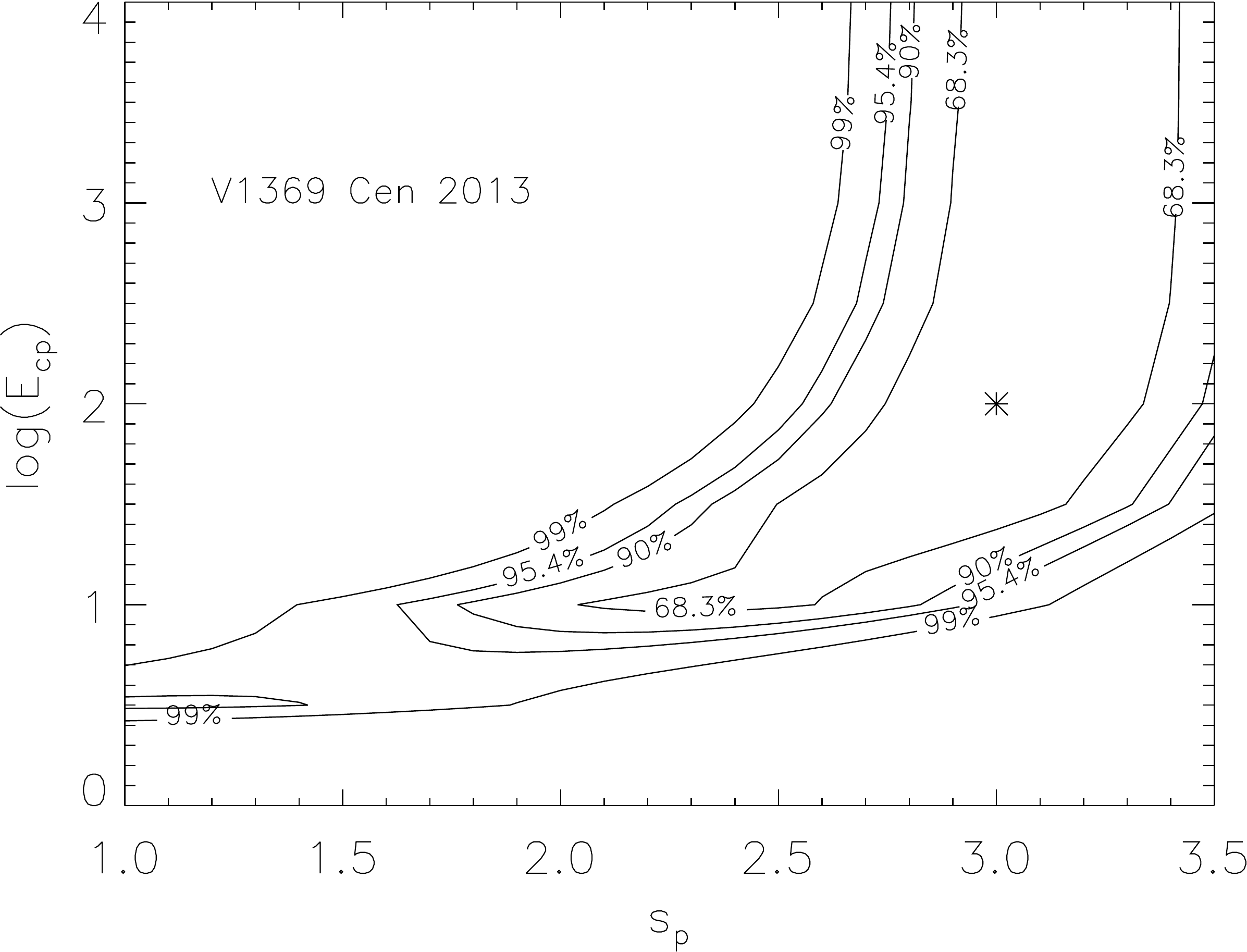}\includegraphics[width=6.1cm,angle=90]{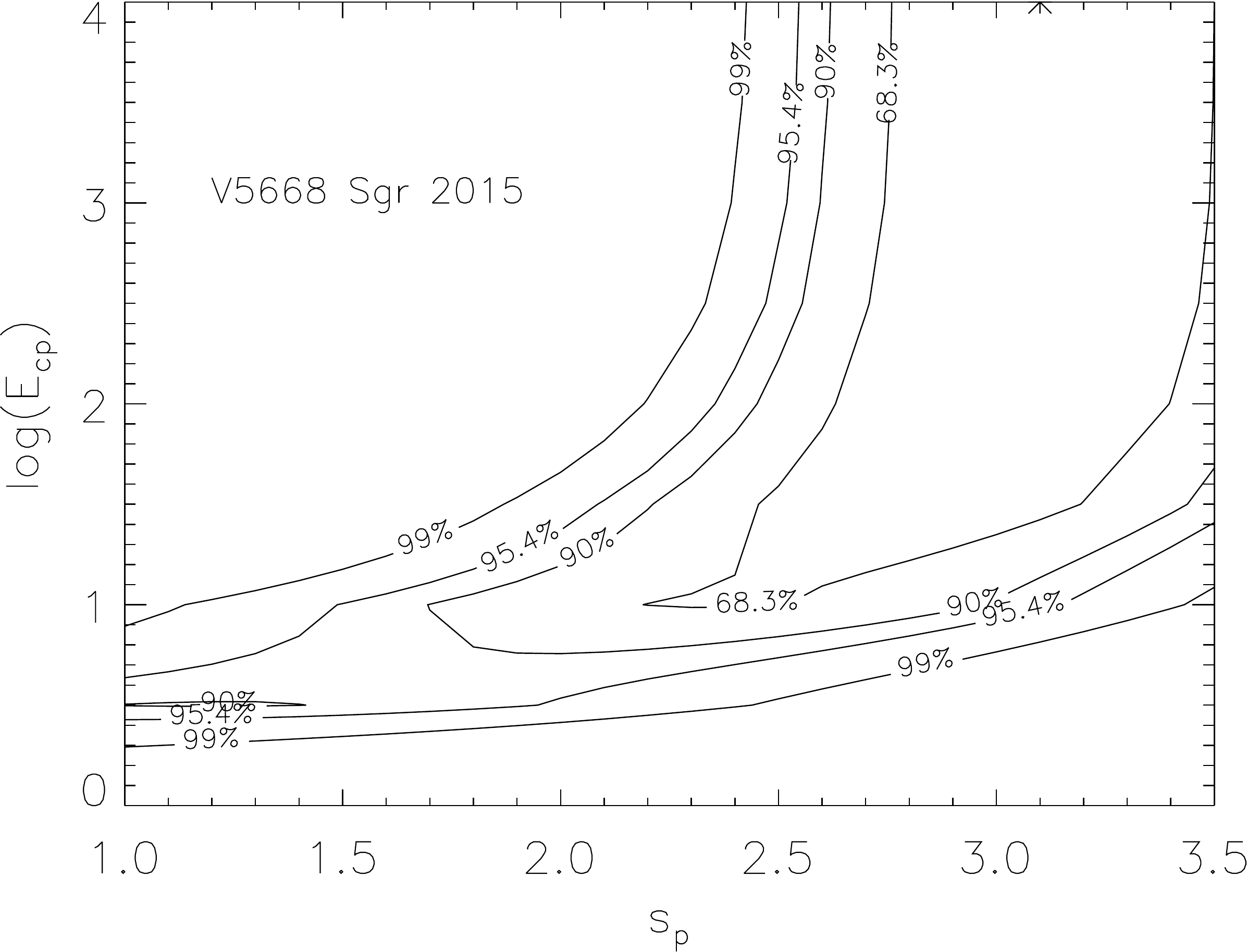}
\includegraphics[width=6cm,angle=90]{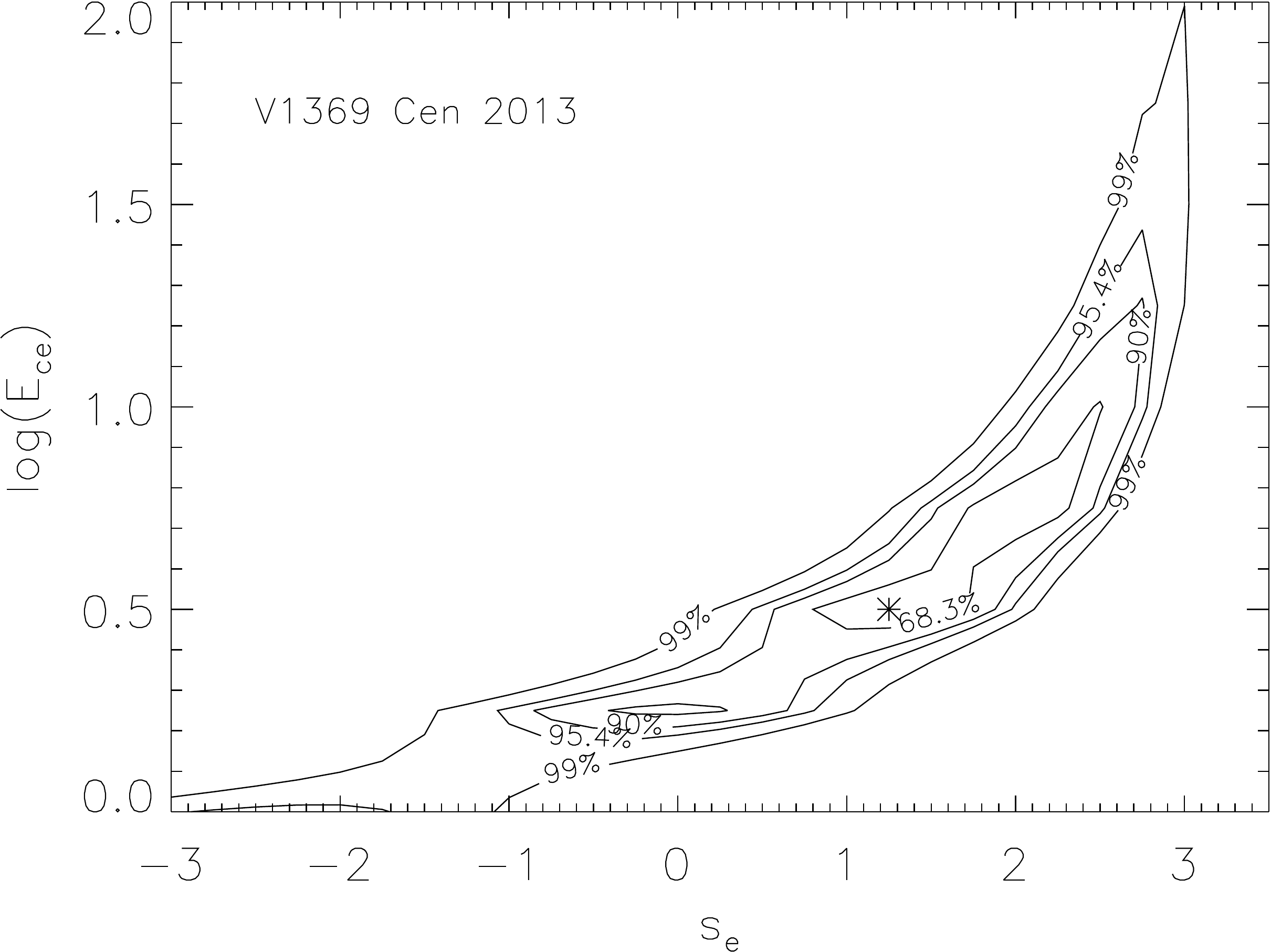}\includegraphics[width=6cm,angle=90]{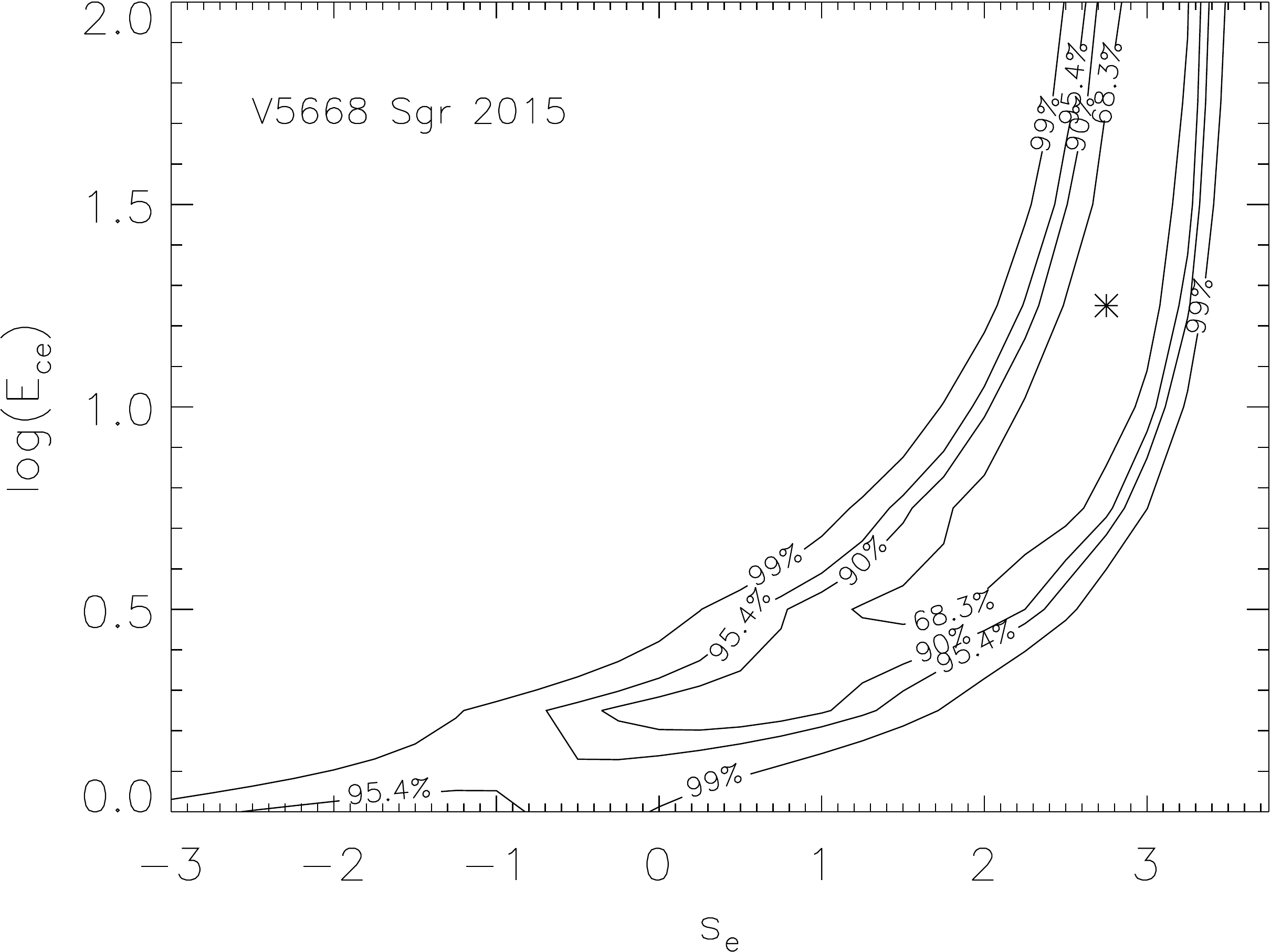}
  \end{center}
\caption{Confidence intervals for the hadronic (top panels) and leptonic (bottom panels) model fits to the \Fermi-LAT data for \ncen\ (left panels) and \nsgr\ (right panels). The best-fit values of the slopes and cutoff energies [GeV] in the hadronic ($s_{\rm p}$, $E_{\rm cp}$) and leptonic ($s_{\rm e}$, $E_{\rm ce}$) models (Table~2) are indicated in each panel with an asterisk.}
\label{figure-4}
\end{figure*}

\subsection{Spectral Analysis, Source Localization, and Modeling}

To derive average LAT spectra, we selected data covering the total durations (39\,days for \ncen\ and  55\,days for \nsgr) determined from the intervals spanned by the TS $\geq 4$ points in the 1-day bin light curves. The spectra were extracted using {\tt gtlike} by fitting the flux from each nova in 12 logarithmic bins from 100\,MeV to 100\,GeV. As in \citet{ack14}, we fit the data assuming a single PL with the integrated $>$100\,MeV flux and the photon index as free parameters, and an exponentially cutoff power law (ECPL) which has the cutoff energy as an additional free parameter to gauge spectral curvature. The best-fit PL and ECPL parameters for both novae are presented in Table~2 with Figure~2 (top panels) showing the best-fit analytic curves overlaid onto the extracted LAT spectra. Note that the TS obtained with the ECPL fit in \ncen\ is only slightly larger than the one obtained with the PL fit, amounting to an improvement of $(\Delta{\rm TS})^{1/2} = 3.3\sigma$, while there is a negligible increase in TS in the \nsgr\ case. In the three previously detected \gray\ classical novae \citep{ack14}, the spectra were similar with broad spectral peaks that cut off at energies $\sim$1$-$4\,GeV and emission observed up to $\sim$6$-$10\,GeV (the highest energy extension was observed in \nsco), and we remark only that the relatively poor statistics in the new cases presented here do not permit a robust comparison.

To confirm positional coincidences of the \gray\ sources with the novae, the LAT positions were determined by running the {\tt gtfindsrc} tool assuming the single PL spectral model, using the same time interval used in the average spectral analyses. The \gray\ sources were well-localized with best-fit LAT $95\%$ localization error radii of 0\fdg10 (\ncen) and 0\fdg14 (\nsgr), consistent with their optical positions (Table~1). The smoothed LAT counts maps centered at each novae over the defined total data intervals is shown in Figure~3.

To help gauge whether the \gray\ emission is due to the decay of neutral pions produced by high-energy proton interactions in the ejecta (hadronic model), or due to inverse Compton and bremsstrahlung of high-energy electrons (leptonic model), the parameters of these physical models were also fit to the LAT data. Assuming a single emission region, the spectral distribution of interacting particles is assumed to be an exponentially cutoff power law defined with a slope, a cutoff energy, and a normalization factor. In the hadronic model, the high-energy protons are expected to be uniformly distributed in the expanding shell where they interact with the nuclei of the ejecta. We followed \citet[][\S~S2, therein]{ack14} for the leptonic model where a typical temperature of 15000\,K with radius $3 \times 10^{12}$\,cm for the optically thick surface (equivalent to the nova photosphere in the optical band) was assumed. High-energy electrons are expected to produce \gray\ photons via bremsstrahlung in collisions with atoms in the ejecta, and the emission spectra were calculated assuming a typical ejecta mass of 10$^{-5}$\,M$_\odot$. As in \citet{ack14}, to simplify our estimates, we adopted a spherical geometry for both novae and the mean radius of the ejecta where high-energy electrons are expected to be accelerated and interact via inverse Compton scattering with photospheric photons is $\sim$$4 \times 10^{14}$\,cm, assuming a typical ejecta velocity of $\sim$2000\,km\,s$^{-1}$ at $\sim$50\,days after explosion. The mean density for the bremsstrahlung emission was obtained by dividing the ejecta mass by its mean volume (averaged over the \gray\ emission duration), assuming it is a shell with a relative thickness of $\Delta R/R$ = 0.4. Note the adopted value  of the ejecta velocity was chosen to be consistent with strict lower limits of $\sim$1500$-$2000\,km\,s$^{-1}$ indicated by the half-widths at zero intensity of the Hydrogen lines in late-time optical spectra \citep[$\simgt$50\,days after optical discovery;][]{sho14,gar15}.

The best-fit parameters for the physical models obtained with {\tt gtlike} along with their TS values are summarized in Table~2 with Figure~2 (bottom panels) showing the model curves compared to the LAT spectra. The slopes and the cutoff energies are not well constrained (see Figure~4 for the likelihood contours) which means that a single power law spectrum can explain the LAT data. With only small differences in TS values ($\Delta$TS $\sim$2$\sigma$ for \ncen), the hadronic and leptonic model-fits were statistically indistinguishable, as they were in the three previously detected classical novae.  The values of the best-fit physical parameters are similar to those obtained from the spectral analyses of the previous classical novae detected by \Fermi-LAT, taking into account their uncertainties \citep[cf.,][supplementary Figures S3 and S4 therein]{ack14}. The similarity of these results suggests a common mechanism is responsible for the  \gray\ emission of classical novae with only the emissivity differing significantly between them.

In the hadronic model, the spectral fits to the LAT data allow us to estimate the total energy in high-energy protons, $\epsilon_{\rm p}$, and the conversion efficiency, $\eta_{\rm p} \equiv \epsilon_{\rm p}/\epsilon_{\rm ej}$, where $\epsilon_{\rm ej}$ is the kinetic energy of the ejecta. Here, the conversion efficiencies were calculated with $\epsilon_{\rm p}$ based on the proton spectrum parameters that best fit the observed \gray\ spectral data, and differs from the cosmic-ray (CR) proton production conversion efficiency which is the ratio of the total energy of the CR protons accelerated at the shock front, $\epsilon_{\rm cr}$, to the kinetic energy of the ejecta (see Appendix B). Estimating the conversion efficiency of the shock-accelerated CR proton production requires an accounting of the propagation and energy loss rate of protons in the expanding ejecta and an assumption of how they were injected (e.g., as a single burst, repeated shock events, at a continuous rate, or through wind-wind interactions). For simplicity, we assume high-energy protons collide with target Hydrogen atoms in the ejecta volume with a density averaged over the source lifetime taken to be the LAT-observed emission duration producing the mean observed \gray\ rate \citep[see the online supplements of][for the details of the calculation method]{abd10,ack14}. Under these conditions, with a typical ejecta mass of 10$^{-5}$\,M$_{\odot}$ and velocity of  2000\,km\,s$^{-1}$, the total energies in protons are $\sim$$3.2 \times 10^{43}$\,erg and $\sim$$2.3 \times 10^{43}$\,erg with $\eta_{\rm p}\sim$$8\%$ and $\sim$6$\%$ for \ncen\ and \nsgr, respectively. Note that $\epsilon_{\rm p}$ and $\eta_{\rm p}$ scale with the inverse of the ejecta mass and with the square of the distance, which we have assumed to be uncertain to $\pm 0.5$\,kpc. These values are slightly larger than (but of the same order of magnitude as) the ones obtained for the previous novae, $\epsilon_{\rm p} \sim$(0.7$-$2)\,$\times 10^{43}$ erg and $\eta_{\rm p} \sim$0.1$-$7$\%$ \citep[see Table S3 of][]{ack14}. From the leptonic model-fits, the total energy in electrons, $\epsilon_{\rm e}$ $\sim$$3.6 \times 10^{41}$\,erg and $\sim$$3.8 \times 10^{41}$\,erg for \ncen\ and \nsgr, respectively, with conversion efficiencies, $\eta_{\rm e}$$\sim$$0.1\%$ for both novae. Here again, these values are similar to the ones obtained for previous novae \citep[see Table S4 of][]{ack14}. Modeling the acceleration and propagation of protons including detailed parameterization of the evolving ejecta \citep[see e.g.,][in the case of V407 Cyg]{mar13} with a more realistic geometry could change these estimates, but these calculations are beyond the scope of this paper.

\section{Discussion}

The two optically bright novae are part of an emergent class of transient Galactic \gray\ sources discovered with the \Fermi-LAT. Together with the first LAT-detected symbiotic-like recurrent nova \ncyg\ and the three classical novae reported in \citet{ack14}, this brings the total to six secure detections thus far. This total does not count the marginal 2$-$3$\sigma$ LAT observations of the recurrent symbiotic-like nova V745 Sco in the 1-day intervals on Feb 6th and 7th \citep{che14}, coincident with the optical peak discovery on 2014 Feb 6.694 \citep{stu14} or the current pending candidates from the LAT Pass 8 analysis of all known optical novae observed since \Fermi's launch \citep{fra15}. The earliest LAT novae detections were serendipitously enabled through the \Fermi-LAT's all-sky monitoring mode, but the most recent examples were enabled by transitioning to deeper ToO observations following their optical discoveries by the amateur community. These two new LAT detections keep the \gray\ novae detection rate over the first $\sim$7\,yrs of the \Fermi\ mission at pace with the early rate of $\sim$1\,yr$^{-1}$, consistent with the estimated rate of Galactic novae at $\simlt$4$-$5\,kpc distances \citep{ack14}. 

The average $>$100\,MeV \gray\ flux of \nsgr, $\sim$$1 \times 10^{-7}$\,\phflux, is the lowest so far for any LAT-detected nova, being $\sim$$2\times$ smaller than for \ncen\ and \ndel, and $\sim$5$-$6$\times$ smaller than the other three LAT-detected novae. The corresponding $>$100\,MeV luminosities are $1\times10^{35}\,(D/{\rm 2.5\,kpc})^{2}$\,erg\,s$^{-1}$ for \ncen\ and $3\times10^{34}\,(D/{\rm 2.0\,kpc})^{2}$\,erg\,s$^{-1}$ for \nsgr. The adopted distances are scaled to \ndel\ \citep{sho14} for \ncen, and the suggested range is $D =$ 1.5$-$2.0\,kpc for \nsgr\ \citep{ban15}. Such small distances are supported by the low extinctions, $E(B-V) \sim$0.10$-$0.14 for \ncen\ \citep{izz13,sho14} and $E(B-V) \sim$0.2$-$0.3 for \nsgr\ \citep{kui15}. The implied $>$100\,MeV \gray\ luminosities are thus, systematically, up to a factor of 10 smaller than the observed range of the previously detected novae of (3$-$9)\,$\times 10^{35}$\,erg\,s$^{-1}$ and suggest more distant analogues will be difficult to detect with \Fermi-LAT.  Overall, these two new LAT detections suggest a wider range of \gray\ luminosities than observed in the first reported \gray\ detections, albeit with the inherent uncertainties in their distance determinations \citep[see][]{che15c}.

Interestingly, \ncen\ and \nsgr\ also appear to have longer \gray\ emission durations than previous examples. Because they are fainter in $\gamma$ rays on average than the previous LAT-detected novae, their light curves are more difficult to characterize, but nevertheless appear sporadic in nature with low-significance 1-day detections extended over moderately longer durations (up to $\sim$1$-$2\,months). Specifically, during times of enhanced LAT exposures due to the ToOs, we found the \gray\ onsets delayed with respect to the first optical peaks by $\sim$2\,days (\S~2.1) and $\sim$5$-$8\,days after their optical discoveries. This delay could indicate the timescale for particle acceleration in the ejecta or \gray\ absorption at early times when the ejecta are most dense \citep{abd10,ack14,cho14,met15}. These explanations could have distinct signatures in early, time-resolved LAT spectra, with e.g., the latter scenario predicting progressively harder \gray\ spectra due to increased transmission of higher-energy photons, but the limited statistics in these cases prevent such an analysis. Moreover, the overall low statistics make it difficult to associate the later individual \gray\ detections with subsequent optical peaks although we note the brighter, second optical peak of 3.3\,mag in \ncen\ appears to be followed by a 2-day long \gray\ brightening (centered at \ts\ $+$12\,days) that is also delayed by $\sim$2\,days (see below). The LAT detections overall span the duration of the multi-peaked optical activity in each nova (\S~1; Figure~1), and indicates future \gray\ novae require dedicated long-term increased LAT-exposures to establish any possible correlations with optical activity \citep[see discussions of previous cases in e.g.,][]{met15}.

\begin{table*}
\begin{center}
\caption{Parameters of classical novae detected by the \Fermi-LAT}
\begin{tabular}{lccccc}
\hline
\hline
Nova 	&  Distance & Duration & Fluence & Number of photons & Total energy \\
        & (kpc)  & (days) &  (photons\,cm$^{-2}$) &  (10$^{45}$) & (10$^{42}$ erg) \\
\hline
V1324 Sco 2012 & 4.5 & 17 & 0.72 $\pm$ 0.04 & 1.77 $\pm$ 0.40 & 1.27 $\pm$ 0.29 \\
V959  Mon 2012 & 3.6 & 22 & 0.79 $\pm$ 0.12 & 1.23 $\pm$ 0.39 & 0.67 $\pm$ 0.21 \\
V339  Del 2013 & 4.2 & 27 & 0.45 $\pm$ 0.07 & 0.96 $\pm$ 0.27 & 0.57 $\pm$ 0.16 \\
V1369 Cen 2013 & 2.5 & 39 & 0.71 $\pm$ 0.14 & 0.54 $\pm$ 0.24 & 0.30 $\pm$ 0.13 \\
V5668 Sgr 2015 & 2.0 & 55 & 0.52 $\pm$ 0.11 & 0.25 $\pm$ 0.14 & 0.12 $\pm$ 0.07 \\
\hline
\end{tabular}
\normalsize
\label{table-3}
\end{center}
{\bf Notes.} 
Durations of the $>$100\,MeV \gray\ emission, fluences, total number of photons, and total energy emitted by classical novae as derived from LAT data analysis. For the three previously detected classical novae, the distances adopted were taken from \citet{ack14}. The 1$\sigma$ uncertainties of the total number of photons and total energies take into account an adopted uncertainty of 0.5\,kpc for all the distances.
\end{table*}

\begin{figure*}
  \begin{center}
\includegraphics[width=8.25cm]{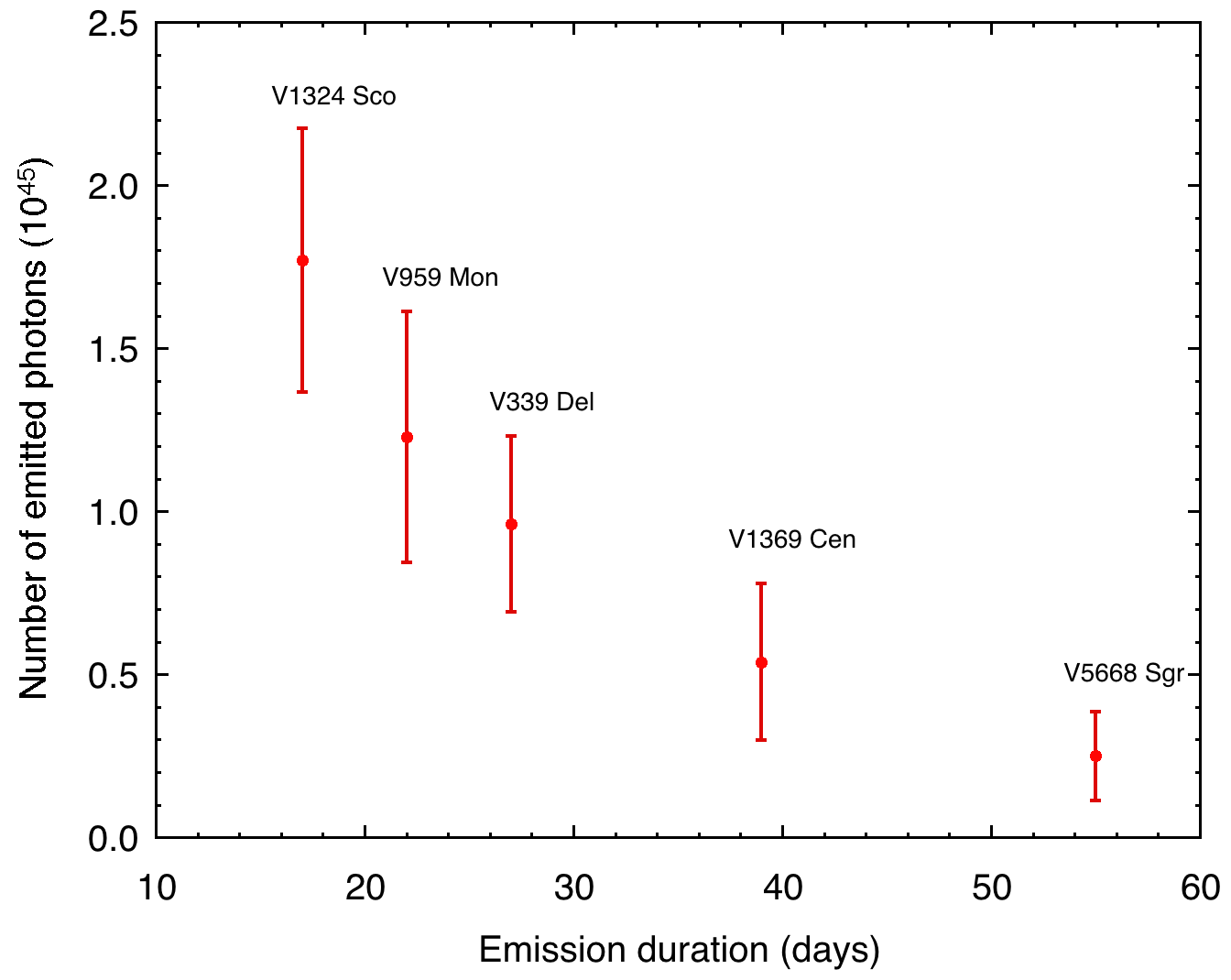}\hspace{-0.05cm}\includegraphics[width=8.25cm]{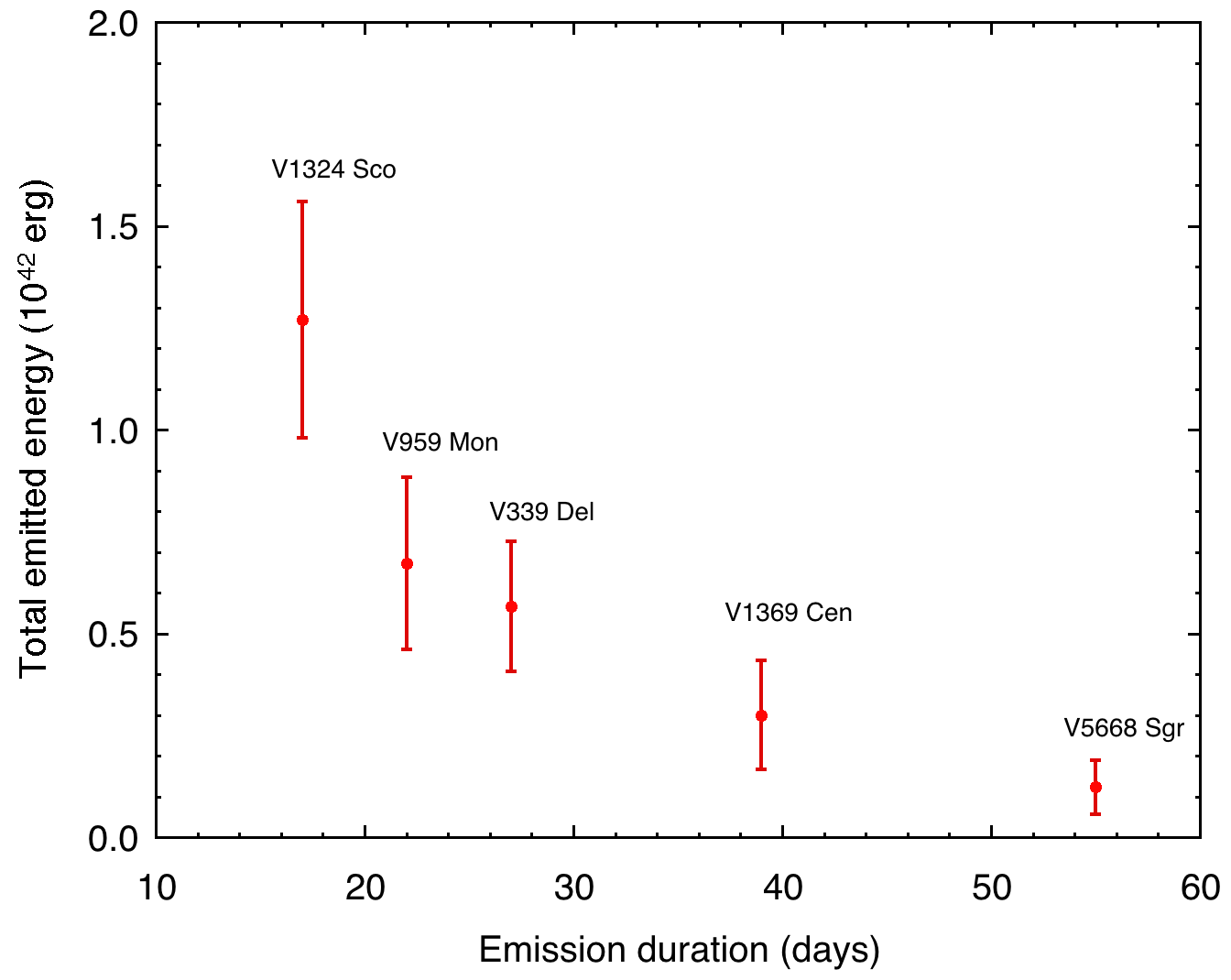}
  \end{center}
\caption{For the five classical novae with reported \Fermi-LAT detections (thus excluding the symbiotic-like recurrent nova \ncyg), the total number of $>$100\,MeV \gray\ photons emitted (left panel) and total emitted energies (right panel) are plotted versus the LAT-measured \gray\ durations.}
\label{figure-5}
\end{figure*}

That the LAT-observed durations of these two classical novae are longer than in the three previously detected examples is of interest, particularly with the lower \gray\ luminosities of the former. We can thus remark on a plausible relation between the total emission durations of the $>$100\,MeV emission found in our LAT-based studies and the mean \gray\ luminosity in the same band (in photons s$^{-1}$). Because the mean luminosity logically decreases with the integration period for transient events, we estimated the total number of $>$100\,MeV photons emitted as well as the total energies over the emission duration and found that the longer the \gray\ duration, the smaller the total number of emitted photons and total energies (Figure~\ref{figure-5}, Table~3). As an intermediate step, the fluences were calculated by multiplying the best-fit average $>$100\,MeV fluxes from the ECPL fits \citep[see Table~2 and Table~S1 of][]{ack14} by the total durations. Note the total energies for the three previously detected classical novae were presented in \citet{ack14} assuming the PL fits, and are consistent with the values presented here. The two new LAT-detected classical novae probe systematically lower values of total energies as the previous examples, as well as total number of emitted photons. Despite the intriguing apparent inverse relationship of these quantities with \gray\ durations, we cannot rule out selection effects related to the \Fermi-LAT sensitivity (e.g., some photons are missing because the flux is too low to be detected) or conclude whether there is a relationship at all due, e.g., to the uncertain distances\footnote{For example, taking the larger distance of $>$6.5\,kpc for \nsco\ \citep{fin15} than the 4.5\,kpc we adopted, the number of emitted photons increases from $1.8 \times 10^{45}$ to $>$$3.8 \times 10^{45}$.}. A more extensive sample of \Fermi-LAT detected novae could test whether this relation is due to physics of the outbursts, particle acceleration, and ejecta parameters -- e.g., for short emission durations, the putative shocks cross more compact ejecta (i.e., with higher density) and for longer emission durations the shock goes through  ejecta with larger radii (with lower density) and is less efficient in producing $\gamma$ rays. 

There has been a long history of predictions (\S~1) for\,MeV line emission promptly after the thermonuclear runaway when the explosion is optically thick, i.e., as the optical brightness peaks and before the GeV emission onset. Specifically, line emission will be expected at 478\,keV and 1275\,keV for CO and ONe novae, respectively, together with positron annihilation line emission at 511\,keV and continuum down to 20$-$30\,keV due to Comptonization and positronium emission. The multiple optical peaks observed in \ncen\ and \nsgr, with apparent concomitant $>$100\,MeV band emission could signal renewed outbursts, originating either from repeated shock events in the ejecta or distinct outbursts from the white dwarf surface. In the former, the \gray\ observations probe the density profile of the ejecta where the shock could interact with different density material resulting in varying rates of particle injection for acceleration processes and/or different target density material for interactions producing $\gamma$ rays.   In the latter case, any accompanying radioactive decay emission will depend on ejecta mass and ejecta velocity for a given white dwarf mass \citep{her02}, which can be different for each subsequent outburst. Previous observations of emission up to 70\,keV with \Suzaku\ \citep{tak09} and the likely $\sim$0.1\,MeV \Compton/OSSE detection of \nvel\ 1999 \citep{che15c} suggest extended timescale emission in classical novae, and should motivate continuum emission searches with \textit{INTEGRAL} \citep{win03} and \textit{ASTRO-H} \citep{tak12,cop14} at later times than so far considered for the nuclear decay emission.

\section{Conclusions}  

The newly reported $>$100\,MeV \gray\ detections of \ncen\ and \nsgr\ add to the four previously detected novae by the \Fermi-LAT.  These recent detections may be revealing a wider diversity in \gray\ properties, being fainter and seemingly characterized by longer total durations.  Further serendipitous and targeted \gray\ detections through the continued \Fermi\ mission, as well as studies of non-LAT detected novae \citep{che12,fra15} could further broaden the range of observed \gray\ properties. While observationally based scenarios have been developed for symbiotic systems \citep{tat07}, and \ncyg\ in particular \citep{abd10,cho12,nel12,orl12,mar13,pan15}, understanding the particle acceleration revealed in the classical novae is still an open problem. Thus future \Fermi-LAT observations of novae (where deep pointed observations are imperative to study the brightest \gray\ emitting novae in detail) can be exploited with multi-wavelength observations and modeling to test scenarios for how, when, and where the putative shocks are generated, e.g., in internal shocks or strong turbulence driven in the ejecta \citep{ack14} or in wind-wind interactions \citep{cho14}, and can implicate the underlying \gray\ emission mechanism \citep[see e.g.,][]{met15}.

At the highest energies, one can anticipate very high energy (VHE; $>$0.1\,TeV) emission as well as a transient neutrino signal in the pion-decay scenario for symbiotic novae \citep{raz10,ali12,bed13} and for classical novae \citep{ahn15,met16}. Amongst the recurrent novae with shorter recurrence times \citep{sch10}, we may anticipate an explosion from RS Oph during the \Fermi\ mission and in the era of the upcoming Cherenkov Telescope Array. Naively scaling by distance, we may expect the even more nearby system, T CrB ($D \sim$0.8\,kpc), to produce an order of magnitude brighter GeV source than V407 Cyg ($D\sim$2.7\,kpc), with emission extending below MeV and up to VHE energies. Any particularly nearby, $D \simlt$\,1$-$2\,kpc, classical nova in the future will provide a useful contrasting case to the symbiotic/recurrent systems. 

With the study of novae now expanded into the \gray\ regime, the words of \citet[][pg.~VII]{pay57} ring even more true today, ``I have become convinced that the {\it whole} nova phenomenon must be studied; the variations of total light and continuum, of radial velocity, and of the intensities and profiles of absorption and emission lines must be seen as connected parts of one physical phenomenon, rather than as isolated data that can be understood separately.'' With the expanding number of LAT-detected novae, existing models are open to a greater level of scrutiny, and the opportunities with \Fermi\ should be pursued in earnest.

\acknowledgments
\vspace{-0.15in}

The \textit{Fermi} LAT Collaboration acknowledges generous ongoing support from a number of agencies and institutes that have supported both the development and the operation of the LAT as well as scientific data analysis. These include the National Aeronautics and Space Administration and the Department of Energy in the United States, the Commissariat \`a l'Energie Atomique and the Centre National de la Recherche Scientifique / Institut National de Physique Nucl\'eaire et de Physique des Particules in France, the Agenzia Spaziale Italiana and the Istituto Nazionale di Fisica Nucleare in Italy, the Ministry of Education, Culture, Sports, Science and Technology (MEXT), High Energy Accelerator Research Organization (KEK) and Japan Aerospace Exploration Agency (JAXA) in Japan, and the K.~A.~Wallenberg Foundation, the Swedish Research Council and the Swedish National Space Board in Sweden.
 
Additional support for science analysis during the operations phase is gratefully acknowledged from the Istituto Nazionale di Astrofisica in Italy and the Centre National d'\'Etudes Spatiales in France.

We thank the \Fermi\ operations team at the Science Support Center for their prompt scheduling of the ToO observations. We acknowledge with thanks the variable star observations from the AAVSO International Database contributed by observers worldwide and used in this research, and the dedicated observers of the ARAS group. We are grateful to G.~Dubus for detailed comments that helped improve the manuscript and the anonymous referee for interesting comments.

C.C.C.~was supported at NRL by a Karles' Fellowship and by NASA through Guest Investigator programs 12-FERMI12-0026 and 13-FERMI13-0008.
\L .S. was supported by Polish NSC grant DEC-2012/04/A/ST9/00083.
S.S.~acknowledges partial support from NSF and NASA grants to ASU.

{\it Facilities:} \facility{Fermi}, \facility{AAVSO}, \facility{ARAS}

{}

\appendix

\section{A. \Fermi-LAT Light Curves}

The \Fermi-LAT 1-day bin light curve data, TS, and exposure values are presented in Figure~A1, with the fluxes and TS values tabulated in Table~A1.

\renewcommand{\thefigure}{A\arabic{figure}}
\setcounter{figure}{0}    

\begin{figure*}[h]
  \begin{center}
    \includegraphics[width=8cm]{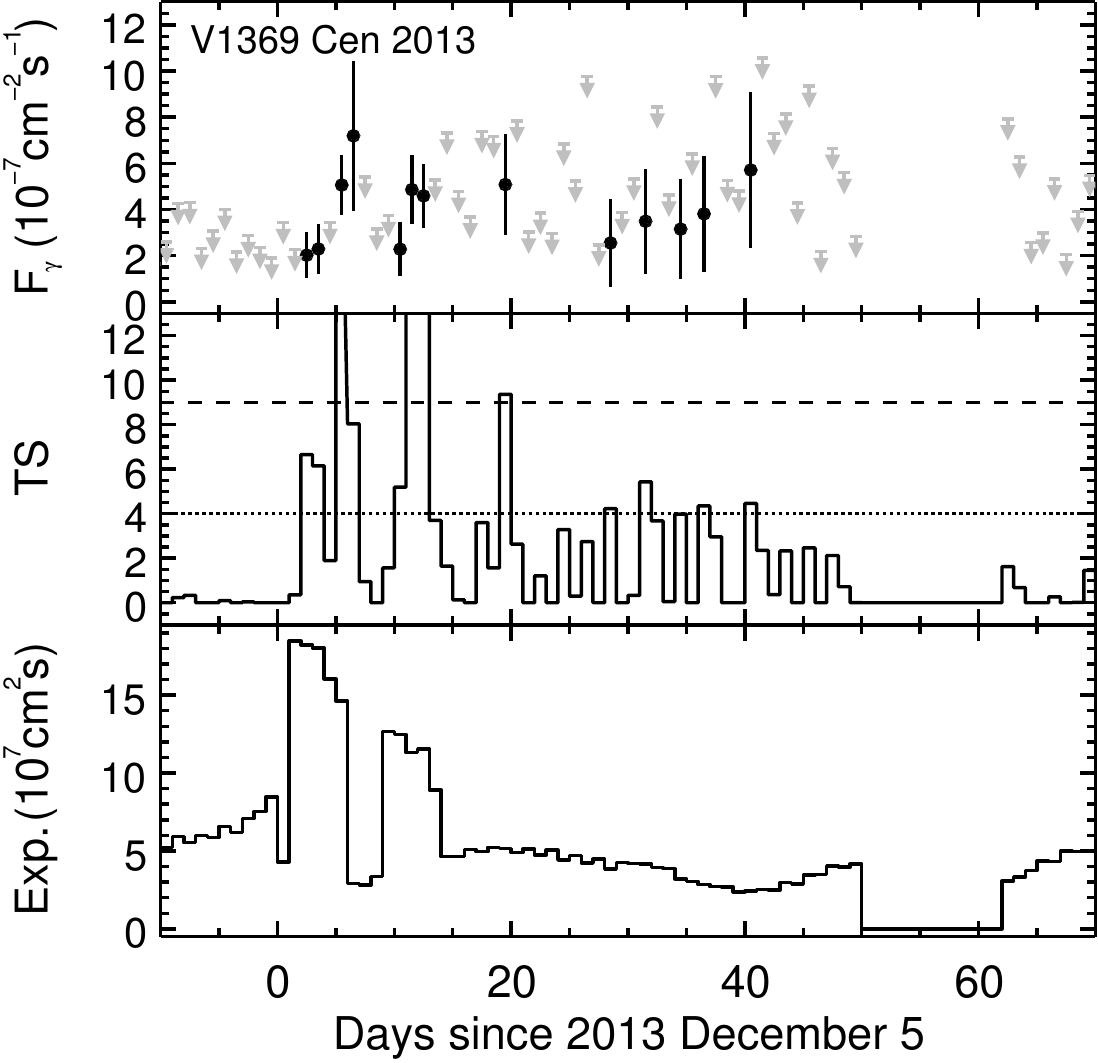}\hspace{.25cm}\includegraphics[width=8cm]{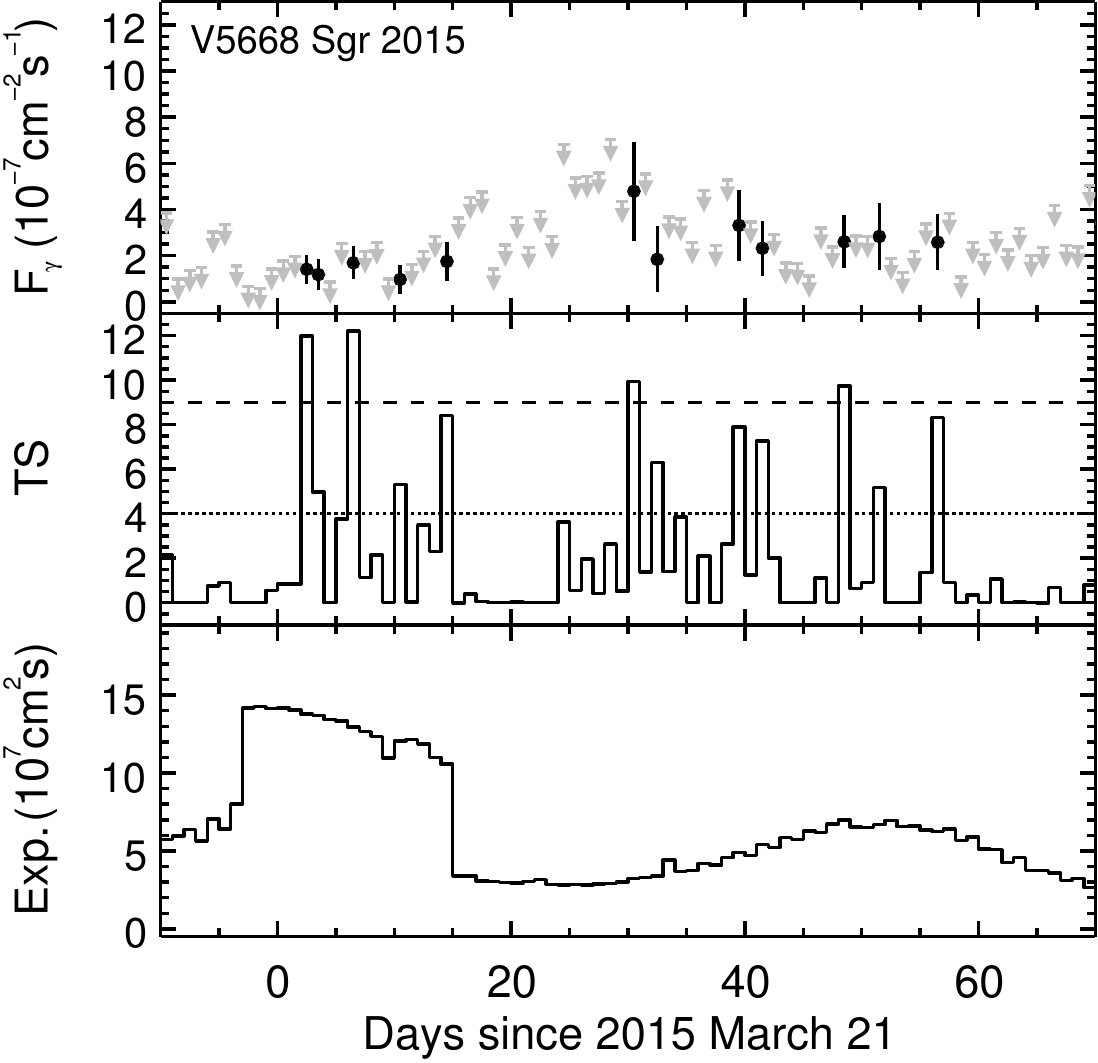}
  \end{center}
\caption{\Fermi-LAT 1-day bin $>$100\,MeV \gray\ data for \ncen\ (left) and \nsgr\ (right). The light curves (top panels) are as presented in Figure~1, along with the corresponding TS (middle panels) and exposure values (bottom panels). The corresponding panels for the two novae show the same range to ease comparison. In the middle panels, horizontal lines indicate values of TS = 4 (dotted) and TS = 9 (dashed). With the range chosen in the middle panel, three larger TS values in \ncen\ were cropped (TS = 37.6, 20.3, and 19.3 centered respectively on \ts $+$5.5, $+$11.5, and $+$12.5\,days).}
\label{figure-appendix}
\end{figure*}

\renewcommand{\thetable}{A\arabic{table}}
\setcounter{table}{0}    

\begin{table*}
\vspace{-0.1in}
\footnotesize
\tiny
  \begin{center}
\caption{\footnotesize \Fermi-LAT $>$100\,MeV \gray\ light curve data}
\vspace{-0.1in}
\begin{tabular}{lcccc}
\hline\hline
\multicolumn{1}{l}{Days since} &
\multicolumn{2}{c}{\ncen\ 2013} &
\multicolumn{2}{c}{\nsgr\ 2015} \\
\multicolumn{1}{l}{start} &
\multicolumn{1}{c}{TS} &
\multicolumn{1}{c}{$F_{\gamma}$} &
\multicolumn{1}{c}{TS} &
\multicolumn{1}{c}{$F_{\gamma}$} \\
\hline
 --9.5	&  0.0  & $< 2.6$        &  2.4	 & $< 4.3$	 \\ 
 --8.5	&  0.2  & $< 4.3$        &  0.0	 & $< 1.1$	 \\ 
 --7.5	&  0.3  & $< 4.3$        &  0.0	 & $< 1.5$	 \\ 
 --6.5	&  0.0  & $< 2.3$        &  0.0	 & $< 1.6$	 \\ 
 --5.5	&  0.0  & $< 3.1$        &  0.8	 & $< 3.4$	 \\ 
 --4.5	&  0.1  & $< 4.0$        &  0.9	 & $< 3.6$	 \\ 
 --3.5	&  0.0  & $< 2.2$        &  0.0	 & $< 1.8$	 \\ 
 --2.5	&  0.0  & $< 2.9$        &  0.0	 & $< 0.7$	 \\ 
 --1.5	&  0.0  & $< 2.4$        &  0.0	 & $< 0.6$	 \\ 
 --0.5	&  0.0  & $< 1.9$        &  0.5	 & $< 1.5$	 \\ 
  0.5	&  0.0  & $< 3.4$        &  0.7	 & $< 1.9$	 \\ 
  1.5	&  0.4  & $< 2.3$        &  0.9	 & $< 2.2$	 \\ 
  2.5	&  6.7  & $ 2.0 \pm 1.0$ & 11.5	 & $ 1.6 \pm 0.7$ \\ 
  3.5	&  6.1  & $ 2.3 \pm 1.1$ &  5.3	 & $ 1.3 \pm 0.7$ \\ 
  4.5	&  1.9  & $< 3.4$        &  0.0	 & $< 1.0$	 \\ 
  5.5	& 37.6  & $ 5.1 \pm 1.3$ &  4.0	 & $ 1.2 \pm 0.7$ \\ 
  6.5	&  8.0  & $ 7.2 \pm 3.2$ & 11.6	 & $ 1.8 \pm 0.8$ \\ 
  7.5	&  0.9  & $< 5.4$        &  1.1	 & $< 2.4$	 \\ 
  8.5	&  0.0  & $< 3.2$        &  2.3	 & $< 2.7$	 \\ 
  9.5	&  1.6  & $< 3.7$        &  0.0	 & $< 1.1$	 \\ 
 10.5	&  5.2  & $ 2.3 \pm 1.2$ &  5.0	 & $ 1.1 \pm 0.7$ \\ 
 11.5	& 20.3  & $ 4.9 \pm 1.5$ &  0.1	 & $< 1.9$	 \\ 
 12.5	& 19.3  & $ 4.6 \pm 1.4$ &  3.1	 & $< 2.3$	 \\ 
 13.5	&  3.7  & $< 5.3$        &  2.8	 & $< 3.1$	 \\ 
 14.5	&  1.6  & $< 7.3$        &  8.5	 & $ 1.9 \pm 0.9$ \\ 
 15.5	&  0.1  & $< 4.8$        &  0.0	 & $< 3.9$	 \\ 
 16.5	&  0.0  & $< 3.7$        &  0.5	 & $< 5.0$	 \\ 
 17.5	&  3.6  & $< 7.4$        &  0.2	 & $< 5.6$	 \\ 
 18.5	&  1.6  & $< 7.2$        &  0.0	 & $< 1.6$	 \\ 
 19.5	&  9.4  & $ 5.1 \pm 2.2$ &  0.0	 & $< 2.7$	 \\ 
 20.5	&  2.6  & $< 7.8$        &  0.1	 & $< 4.0$	 \\ 
 21.5	&  0.0  & $< 3.0$        &  0.0	 & $< 2.7$	 \\ 
 22.5	&  1.2  & $< 3.8$        &  0.1	 & $< 4.6$	 \\ 
 23.5	&  0.0  & $< 3.0$        &  0.0	 & $< 3.3$	 \\ 
 24.5	&  3.3  & $< 6.8$        &  3.4	 & $< 7.3$	 \\ 
 25.5	&  0.3  & $< 5.2$        &  0.7	 & $< 6.0$	 \\ 
 26.5	&  2.7  & $< 9.8$        &  1.8	 & $< 5.8$	 \\ 
 27.5	&  0.0  & $< 2.5$        &  0.6	 & $< 6.3$	 \\ 
 28.5	&  4.2  & $ 2.5 \pm 1.9$ &  2.9	 & $< 7.5$	 \\ 
 29.5	&  0.0  & $< 3.9$        &  0.5	 & $< 4.7$	 \\ 
 30.5	&  0.3  & $< 5.3$        & 10.2	 & $ 5.3 \pm 2.3$ \\ 
 31.5	&  5.4  & $ 3.5 \pm 2.3$ &  1.6	 & $< 6.2$	 \\ 
 32.5	&  3.7  & $< 8.4$        &  6.0	 & $ 2.1 \pm 1.6$ \\ 
 33.5	&  0.1  & $< 4.6$        &  1.4	 & $< 4.0$	 \\ 
 34.5	&  4.0  & $ 3.1 \pm 2.2$ &  3.4	 & $< 3.9$	 \\ 
 35.5	&  0.0  & $< 6.4$        &  0.0	 & $< 3.0$	 \\ 
 36.5	&  4.3  & $ 3.8 \pm 2.5$ &  2.3	 & $< 5.3$	 \\ 
 37.5	&  3.0  & $< 9.8$        &  0.0	 & $< 2.7$	 \\ 
 38.5	&  0.0  & $< 5.2$        &  2.9	 & $< 5.8$	 \\ 
 39.5	&  0.0  & $< 4.8$        &  8.0	 & $ 3.6 \pm 1.6$ \\ 
 40.5	&  4.5  & $ 5.7 \pm 3.4$ &  1.4	 & $< 3.9$	 \\ 
 41.5	&  2.4  & $<10.6$        &  7.1	 & $ 2.5 \pm 1.3$ \\ 
 42.5	&  0.4  & $< 7.3$        &  1.8	 & $< 3.1$	 \\ 
 43.5	&  2.3  & $< 8.1$        &  0.0	 & $< 2.0$	 \\ 
 44.5	&  0.0  & $< 4.3$        &  0.0	 & $< 2.0$	 \\ 
 45.5	&  2.5  & $< 9.3$        &  0.0	 & $< 1.3$	 \\ 
 46.5	&  0.0  & $< 2.2$        &  1.2	 & $< 3.5$	 \\ 
 47.5	&  2.1  & $< 6.7$        &  0.0	 & $< 2.0$	 \\ 
 48.5	&  0.7  & $< 5.6$        &  9.7	 & $ 2.8 \pm 1.2$ \\ 
 49.5	&  0.0  & $< 2.8$        &  0.8	 & $< 3.2$	 \\ 
 50.5	&  -	& -	  	 &  0.9	 & $< 3.1$	 \\ 
 51.5	&  -	& -	  	 &  6.1	 & $ 3.2 \pm 1.5$ \\ 
 52.5	&  -	& -	  	 &  0.0	 & $< 2.1$	 \\ 
 53.5	&  -	& -	  	 &  0.0	 & $< 1.4$	 \\ 
 54.5	&  -	& -	  	 &  0.0	 & $< 2.6$	 \\ 
 55.5	&  -	& -	  	 &  1.6	 & $< 3.8$	 \\ 
 56.5	&  -	& -	  	 &  8.6	 & $ 2.9 \pm 1.3$ \\ 
 57.5	&  -	& -	  	 &  1.4	 & $< 4.3$	 \\ 
 58.5	&  -	& -	  	 &  0.0	 & $< 1.2$	 \\ 
 59.5	&  -	& -	  	 &  0.4	 & $< 2.9$	 \\ 
 60.5	&  -	& -	  	 &  0.0	 & $< 2.4$	 \\ 
 61.5	&  -	& -	  	 &  1.0	 & $< 3.2$	 \\ 
 62.5	&  1.6  & $< 7.9$	 &  0.0	 & $< 2.5$	 \\ 
 63.5	&  0.7  & $< 6.3$	 &  0.0	 & $< 3.5$	 \\ 
 64.5	&  0.0  & $< 2.6$	 &  0.0	 & $< 2.3$	 \\ 
 65.5	&  0.0  & $< 3.0$	 &  0.0	 & $< 2.6$	 \\ 
 66.5	&  0.3  & $< 5.3$	 &  0.7	 & $< 4.5$	 \\ 
 67.5	&  0.0  & $< 2.0$	 &  0.0	 & $< 2.7$	 \\ 
 68.5	&  0.0  & $< 3.9$	 &  0.0	 & $< 2.6$	 \\ 
 69.5	&  1.5  & $< 5.5$	 &  0.8	 & $< 5.5$	 \\ 
\hline
\end{tabular}
  \end{center} 
\vspace{-0.1in}
\footnotesize \small
{\bf Notes.} 
Days indicated are the centers of the 1-day bins relative to the defined start times (\ts). TS values and $>$100\,MeV fluxes, $F_{\gamma}$, in units of $10^{-7}$\,\phflux\ ($2\sigma$ upper limits when TS $<4$) are presented. The gap in observations of \ncen\ from \ts $+$50.0 to $+$62.0\,days is due to a LAT ToO on SN2014J in M82.
\label{table-lc}
\end{table*}

\newpage
\clearpage

\section{B. Cosmic Ray Proton Energetics}

Assuming that cosmic ray (CR) protons are injected at the shock front of a nova explosion at a constant rate with an energy spectrum $\dot{Q}(E)$, such that the total energy stored in CR protons over the source lifetime of the ejecta, $\tau_{\rm ej}$, is: 
\begin{equation}
\epsilon_{\rm cr} = \int_0^{\tau_{\rm ej}}\!\!\!dt \,\, \int \!\!dE \, E \, \dot{Q}(E) = \tau_{\rm ej} \times \int dE \, E \, \dot{Q}(E) \equiv \tau_{\rm ej} \times \dot{Q}_{\rm cr} \, .
\end{equation}
We define the CR proton production efficiency factor:
\begin{equation}
\eta_{\rm cr} \equiv \frac{\epsilon_{\rm cr}}{\epsilon_{\rm ej}} \, ,
\end{equation}
where $\epsilon_{\rm ej}$ is the total kinetic energy of the ejecta, hence:
\begin{equation}
\dot{Q}_{\rm cr} = \frac{\eta_{\rm cr} \, \epsilon_{\rm ej}}{\tau_{\rm ej}} \, .
\end{equation}

CR protons injected at the shock front are advected downstream, where they interact with the ambient plasma through proton-proton ($pp$) interactions, and may eventually escape diffusively. The characteristic timescales for these processes are $\tau_{\rm pp}$ and $\tau_{\rm esc}$, respectively. Therefore, the CR proton energy spectrum, $N(E,t)$, evolves accordingly to:
\begin{equation}
\frac{\partial N(E,t)}{\partial t} + \frac{N(E,t)}{\tau_{\rm min}} =  \dot{Q}(E) \, ,
\end{equation}
where $\tau_{\rm min} \equiv \min(\tau_{\rm pp}, \tau_{\rm esc})$. Note here $pp$ interactions are modelled as ``catastrophic'' energy losses, and CR proton diffusion in the downstream is approximated by a simple ``escape'' term. The solution to the above equation at a given moment of time $t=\tau_{\rm ej}$ reads then as:
\begin{equation}
N(E,\tau_{\rm ej}) = \tau_{\rm min} \, \left(1-e^{-\tau_{\rm ej}/\tau_{\rm min}}\right) \times \dot{Q}(E)  \, ,
\end{equation}
assuming that $\tau_{\rm min}$ is independent of time.

The $pp$ interactions lead to the production of $\gamma$-ray photons with energies $\varepsilon \simeq 0.2 \, E$ \citep[see][]{abd10}, with a monochromatic luminosity at a given moment $t=\tau_{\rm ej}$  approximately:
\begin{equation}
\left[\varepsilon L_{\varepsilon}\right] \simeq \eta_{\gamma} \,\, \frac{E^2 N(E,\tau_{\rm ej})}{\tau_{\rm pp}} \, ,
\end{equation}
where the numerical factor $\eta_{\gamma}$ is of the order of $\sim 0.1$ ($\eta_{\gamma} <1$ because, in addition to $\gamma$ rays, secondary $e\pm$ pairs and neutrinos are also produced in $pp$ interactions). Hence, around the peak photon energies for which $\left[\varepsilon L_{\varepsilon}\right] \simeq L_{\gamma}$, with corresponding $E^2 \dot{Q}(E) \simeq \dot{Q}_{\rm cr}$, one has:
\begin{equation}
L_{\gamma} \sim \eta_{\gamma} \, \eta_{\rm cr} \, \epsilon_{\rm ej} \,\, \frac{\tau_{\rm min}}{\tau_{\rm ej} \, \tau_{\rm pp}} \, \, \left(1-e^{-\tau_{\rm ej}/\tau_{\rm min}}\right) 
\end{equation}
(equations B3, B5, and B6).

Let us now consider the three possible cases:

\begin{itemize}
\item[i)]  $\tau_{\rm ej} \ll \tau_{\rm min}$, where the source lifetime is much shorter than either the $pp$ interaction timescale or the CR proton escape timescale. Under such conditions, $\left(1-e^{-\tau_{\rm ej}/\tau_{\rm min}}\right) \sim \tau_{\rm ej}/\tau_{\rm min}$, thus:
\begin{equation}
L_{\gamma} \sim \eta_{\gamma} \, \eta_{\rm cr} \,\, \frac{\epsilon_{\rm ej}}{\tau_{\rm pp}} \, .
\end{equation}
This case describes a young system where CR protons injected downstream have yet to escape, and only \emph{some} fraction ($\tau_{\rm ej}/\tau_{\rm pp} \ll 1$) of the total energy stored in CR protons at the shock front could be transferred to the $\gamma$-ray emission (depending on $\tau_{\rm pp}$ and regardless of $\tau_{\rm esc}$).

\item[ii)] $\tau_{\rm pp} \ll \tau_{\rm esc}$ and $\tau_{\rm pp} \ll \tau_{\rm ej}$, where the $pp$ interaction timescale is much shorter than both the CR proton escape timescale and the source lifetime. Under such conditions, $\tau_{\rm min} = \tau_{\rm pp}$ and $\left(1-e^{-\tau_{\rm ej}/\tau_{\rm pp}}\right) \sim 1$, thus:
\begin{equation}
L_{\gamma} \sim \eta_{\gamma} \, \eta_{\rm cr} \,\, \frac{\epsilon_{\rm ej}}{\tau_{\rm ej}} \, .
\end{equation}
Here, CR protons injected downstream produce $\gamma$-rays so efficiently that all the energy stored in CR protons at the shock front is transferred immediately to $\gamma$-ray emission.

\item[iii)]  $\tau_{\rm esc} \ll \tau_{\rm pp}$ and $\tau_{\rm esc} \ll \tau_{\rm ej}$, where the escape timescale is much shorter than both the $pp$ interaction timescale and the source lifetime. Under such conditions, $\tau_{\rm min} = \tau_{\rm esc}$ and $\left(1-e^{-\tau_{\rm ej}/\tau_{\rm esc}}\right) \sim 1$, thus:
\begin{equation}
L_{\gamma} \sim \eta_{\gamma} \, \eta_{\rm cr} \,\, \frac{\epsilon_{\rm ej}}{\tau_{\rm ej}} \, \frac{\tau_{\rm esc}}{\tau_{\rm pp}} \, .
\end{equation}
Here, CR protons injected downstream escape so efficiently, that only a very small fraction ($\tau_{\rm esc}/\tau_{\rm pp} \ll 1$) of the total energy stored in CR protons at the shock front is transferred to the $\gamma$-ray emission over the source lifetime.
\end{itemize}

For the hadronic scenario considered in the main text, the condition $\tau_{\rm ej} \ll \tau_{\rm min}$ described in case \#i is applicable.


\begin{thebibliography}{}

\bibitem[Abdo et al.(2010)]{abd10} Abdo, A.~A., Ackermann, M., Ajello, M., et al.\ 2010, Science, 329, 817  

\bibitem[Acero et al.(2015)]{ace15} Acero, F.,  Ackermann, M., Ajello, M., et al.\  2015, \apjs, 218, 23  

\bibitem[Ackermann et al.(2012)]{ack12} Ackermann, M., Ajello, M., Albert, A., et al.\ 2012, \apjs, 203, 4  

\bibitem[Ackermann et al.(2014)]{ack14} Ackermann, M., Ajello, M., Albert, A., et al.\ 2014, Science, 345, 554  

\bibitem[Ahnen et al.(2015)]{ahn15} Ahnen, M.~L., Ansoldi, S., Antonelli, L.~A., et al.\ 2015, \aap, 582, A67 

\bibitem[Aliu et al.(2012)]{ali12} Aliu, E., Archambault, S., Arlen, T., et al.\ 2012, \apj, 754, 77 

\bibitem[Atwood et al.(2009)]{atw09} Atwood, W.~B., Abdo, A.~A., Ackermann, M., et al.\ 2009, \apj, 697, 1071  

\bibitem[Banerjee et al.(2015)]{ban15} Banerjee, D.~P.~K., Srivastava, M.~K., Ashok, N.~M., Venkataraman, V.\ 2015, \mnras, 455, L109

\bibitem[Bednarek(2013)]{bed13} Bednarek, W.\ 2013, Astropart.~Phys., 43, 81

\bibitem[Bode \& Evans(2008)]{bod08} Bode, M.~F., \& Evans, A.\ (Eds.) 2008, Classical Novae, 2nd Edition. Cambridge Astrophysics Series, No.~43, Cambridge: Cambridge University Press

\bibitem[Cheung(2012)]{che12} Cheung, C.~C.\ 2012, in 4th Fermi Symposium, Monterey CA, eConf C121028, 106; arXiv:1304.3475 

\bibitem[Cheung et al.(2010)]{che10} Cheung, C.~C., Donato, D., Wallace, E., et al.\ 2010, \atel, 2487, 1 

\bibitem[Cheung et al.(2013a)]{che13a} Cheung, C.~C., Jean, P., Shore, S.~N.\ 2013a, \atel, 5649, 1

\bibitem[Cheung et al.(2013b)]{che13b} Cheung, C.~C., Jean, P., Shore, S.~N.\ 2013b, \atel, 5653, 1

\bibitem[Cheung et al.(2014)]{che14} Cheung, C.~C., Jean, P., Shore, S.~N.\ 2014, \atel, 5879, 1

\bibitem[Cheung et al.(2015a)]{che15a} Cheung, C.~C., Jean, P., Shore, S.~N.\ 2015a, \atel, 7283, 1

\bibitem[Cheung et al.(2015b)]{che15b} Cheung, C.~C., Jean, P., Shore, S.~N.\ 2015b, \atel, 7315, 1

\bibitem[Cheung et al.(2015c)]{che15c} Cheung, C.~C., Jean, P., Shore, S.~N., Grove, J.~E., Leising, M.\ 2015c, Proceedings of Science (ICRC2015), 34th ICRC, 880; arXiv:1605.01375

\bibitem[Chomiuk et al.(2012)]{cho12} Chomiuk, L., Krauss, M.~I., Rupen, M.~P., et al.\ 2012, \apj, 761, 173 

\bibitem[Chomiuk et al.(2014)]{cho14} Chomiuk, L., Linford, J.~D., Yang, J., et al.\ 2014, \nat, 514, 339 

\bibitem[Clayton(1981)]{cla81} Clayton, D.~D.\ 1981, \apjl, 244, L97 

\bibitem[Clayton \& Hoyle(1974)]{cla74} Clayton, D.~D., \& Hoyle, F.\ 1974, \apjl, 187, L101 

\bibitem[Coppi et al.(2014)]{cop14} Coppi, P., Stawarz, \L., Done, C., et al.\ 2014, \textit{ASTRO-H} Space X-ray Observatory White Paper: broad-band spectroscopy and polarimetry, arXiv:1412.1190 

\bibitem[Finzell et al.(2015)]{fin15} Finzell, T., Chomiuk, L., Munari, U., \& Walter, F.~M.\ 2015, \apj, 809, 160 

\bibitem[Franckowiak et al.(2015)]{fra15} Franckowiak, A., Buson, S., Jean, P., Cheung, T.\ 2015, Abstracts of the Sixth Fermi Symposium, 3179

\bibitem[Garde \& Buil(2015)]{gar15} Garde, O., \&  Buil, C.\ 2015, ARAS Eruptive Stars Information Letter no.~17, F.~Teyssier (Ed.), 2015-05, pp.~11-13

\bibitem[Gehrz et al.(1998)]{geh98} Gehrz, R.~D., Truran, J.~W., Williams, R.~E., \& Starrfield, S.\ 1998, \pasp, 110, 3 

\bibitem[Hernanz(2008)]{her08} Hernanz, M.\ 2008, in Classical Novae, M.~F.~Bode, A.~Evans, Eds.\ (Cambridge: Cambridge University Press), 2nd Ed., 252

\bibitem[Hernanz(2014)]{her14} Hernanz, M.\ 2014, ASPC, 490, 319 

\bibitem[Hernanz et al.(2002)]{her02} Hernanz, M., G{\'o}mez-Gomar, J., \& Jos{\'e}, J.\ 2002, \nar, 46, 559 

\bibitem[Izzo et al.(2013)]{izz13} Izzo, L., Mason, E., Vanzi, L., et al.\ 2013, \atel, 5639, 1

\bibitem[Kornoch(2013)]{kor13} Kornoch, K.\ 2013, \atel, 5621, 1

\bibitem[Kraft(1964)]{kra64} Kraft, R.~P.\ 1964, \apj, 139, 457 

\bibitem[Kuin et al.(2015)]{kui15} Kuin, P., Page, K., Osborne, J., Shore, S., Schwartz, G., Walter, F.\ 2015, \atel, 8275, 1

\bibitem[Lundmark(1921)]{lun21} Lundmark, K.\ 1921, \pasp, 33, 225 

\bibitem[Martin \& Dubus(2013)]{mar13} Martin, P., \& Dubus, G.\ 2013, \aap, 551, A37 

\bibitem[Mattox et al.(1996)]{mat96} Mattox, J.~R., Bertsch, D.~L., Chiang, J., et al.\ 1996, \apj, 461, 396 

\bibitem[McLaughlin(1939)]{mcl39} McLaughlin, D.~B.\ 1939, Popular Astronomy, 47, 410 

\bibitem[Metzger et al.(2015)]{met15} Metzger, B.~D., Finzell, T., Vurm, I., et al.\ 2015, \mnras, 450, 2739  

\bibitem[Metzger et al.(2016)]{met16} Metzger, B.~D., Caprioli, D., Vurm, I., et al.\ 2016, \mnras, 457, 1786

\bibitem[Munari et al.(2015)]{mun15} Munari, U., Maitan, A., Moretti, S., \& Tomaselli, S.\ 2015, \na, 40, 28 

\bibitem[Nelson et al.(2012)]{nel12} Nelson, T., Donato, D., Mukai, K., Sokoloski, J., \& Chomiuk, L.\ 2012, \apj, 748, 43 

\bibitem[Orlando \& Drake(2012)]{orl12} Orlando, S., \& Drake, J.~J.\ 2012, \mnras, 419, 2329 

\bibitem[Pan et al.(2015)]{pan15} Pan, K.-C., Ricker, P.~M., Taam, R.~E.\ 2015, \apj, 806, 27 

\bibitem[Payne-Gaposchkin(1957)]{pay57} Payne-Gaposchkin, C.\ 1957, The Galactic Novae, Amsterdam: North-Holland

\bibitem[Razzaque et al.(2010)]{raz10} Razzaque, S., Jean, P., Mena, O.\ 2010, \prd, 82, 123012

\bibitem[Schaefer(2010)]{sch10} Schaefer, B.~E.\ 2010, \apjs, 187, 275 

\bibitem[Seach et al.(2013)]{sea13} Seach, J., Guido, E., Howes, N., et al.\ 2013, CBET, No.~3732

\bibitem[Seach et al.(2015)]{sea15} Seach, J., Itagaki, K., Guido, E., et al.\ 2015, CBET, No.~4080

\bibitem[Shore et al.(2014)]{sho14} Shore, S.~N., Schwartz, G.~J., Walter, F.~M., et al.\ 2014, \atel, 6413, 1

\bibitem[Starrfield et al.(1978)]{sta78} Starrfield, S., Truran, J.~W., Sparks, W.~M., \& Arnould, M.\ 1978, \apj, 222, 600 

\bibitem[Stubbings \& Pearce(2014)]{stu14} Stubbings, R., Pearce, A.\ 2014, reported by E.~O.~Waagen, CBET, No.~3803

\bibitem[Takahashi et al.(2012)]{tak12} Takahashi, T., Mitsuda, K., Kelley, R., et al.\ 2012, \procspie, 8443, 84431Z

\bibitem[Takei et al.(2009)]{tak09} Takei, D., Tsujimoto, M., Kitamoto, S., et al.\ 2009, \apjl, 697, L54 

\bibitem[Tatischeff \& Hernanz(2007)]{tat07} Tatischeff, V., \& Hernanz, M.\ 2007, \apjl, 663, L101 

\bibitem[Waagen(2015)]{waa15} Waagen, E.~O.\ (Ed.) 2015, AAVSO Alert Notice 519

\bibitem[Williams et al.(2015)]{wil15} Williams, S.~C., Darnley, M.~J., Bode, M.~F.\ 2015, \atel, 7230, 1

\bibitem[Winkler et al.(2003)]{win03} Winkler, C., Courvoisier, T.~J.-L., Di Cocco, G., et al.\ 2003, \aap, 411, L1 

\bibitem[Woudt \& Ribeiro(2014)]{wou14} Woudt, P.~A., \& Ribeiro, V.~A.~R.~M.\ (Eds.) 2014, Stella Novae: Past and Future Decades, ASP Conf.~Ser.~490  

\end{thebibliography}
\end{document}